\documentclass[12pt]{style}
\usepackage{graphicx}
\usepackage{amsmath,amssymb}
\usepackage{bm}

\newcommand{\gras}[1]{\boldsymbol{#1}}

\begin{document}

\title{Isospin effects in $N\approx Z$ nuclei in extended Density Functional Theory}

\author{
Wojciech Satu{\l}a\email{satula@fuw.edu.pl}\\
\it Institute of Theoretical Physics, Faculty of Physics \\
\it University of Warsaw, PL-02-093 Warsaw, Poland\\
Witold Nazarewicz\email{witek@frib.msu.edu }\\
\it Dept. of Physics and Astronomy and FRIB Laboratory\\
\it  Michigan State University, East Lansing, MI 48824, USA \\
\it Institute of Theoretical Physics, Faculty of Physics \\
\it University of Warsaw, PL-02-093 Warsaw, Poland\\
}


\date{}

\maketitle

\begin{abstract}
This paper overviews
various phenomena related to the concept of isospin symmetry. The focus is on $N\approx Z$ nuclei, which are excellent laboratories of  isospin physics. The theoretical framework applied is nuclear Density Functional Theory  and its
 isospin- and angular-momentum projected extensions, as well as symmetry-projected multi-reference models. The  topics covered include:
isospin impurities, superallowed
beta decays,  beta-transitions in mirror nuclei, isospin breaking hadronic interactions, mirror and triplet binding energy differences, and
isoscalar pairing.
\end{abstract}


\section{Introduction}
\label{Sect.01}

Isobaric spin was introduced by Heisenberg in 1932 \cite{[Hei32a]} in order to explain the neutron-proton symmetry.
Subsequently, the isospin symmetry has been widely used in theoretical modeling of the atomic nuclei \cite{[Wig37],[Rob66],[Wil69a]}. Even though
it is broken by the electroweak force, isospin is extremely useful for our understanding of  nuclear structure and decays.

Of particular interest for isospin physics are $N\approx Z$ nuclei. Beyond $^{40}$Ca, many of them
are located far from the
line of beta stability, in close proximity to
the proton drip line. Because of the similarity of proton and neutron shell-model orbits, the $N\approx Z$ systems exhibit unique phenomena related to the attractive nature of the isoscalar component of the nuclear force. Examples are: superallowed Fermi beta decays \cite{[Har15]}, superallowed Gamow-Teller decays \cite{[Hin12]}, superallowed alpha decays \cite{[Lid06]}, Wigner energy \cite{[Sat97]}, isoscalar pairing \cite{[Fra14]}, and collective modes \cite{[Nak94a]}.

The atomic nuclei  with enhanced sensitivity to fundamental symmetries are unique laboratories to search for signals of  new physics beyond the Standard Model.
The $N\approx Z$ nuclei are particularly interesting probes as
the superallowed $I$=0$^+$$\rightarrow$$I$=0$^+$ beta-decays among the isobaric analogue states in the isospin triplet
allow stringent tests of  the conserved vector current hypothesis and  provide  precise values of the strength of the weak
force and  the leading $V_{{\rm ud}}$ element of the Cabbibo-Kobayashi-Maskawa matrix.
Other examples include
Fermi- and Gamow-Teller ground-state beta-decays in $T$=1/2 mirror nuclei, which offer alternative tests of the electroweak sector. In all these cases, high-fidelity theoretical calculations  of radiative corrections and isospin symmetry breaking (ISB) effects are needed
to extract the crucial information from precise measurements.

The theory roadmap for this area involves  ab-initio and configuration interaction (shell model) approaches, and nuclear density functional theory (DFT). The latter is the tool of choice for open-shell, deformed complex systems. The focus of  this paper is on DFT-based frameworks and their extensions, including
multi-reference approaches involving symmetry restoration.
Such models are able to capture core-polarization effects originating from a
subtle interplay between the  long-range Coulomb interaction and short-range hadronic inter-nucleon forces; hence, they
are particularly useful to study various aspects of   electroweak interaction.

The aim of this overview is to present selected aspects of isospin physics in $N\sim Z$ nuclei  in terms of theoretical approaches rooted in nuclear density functional theory.
We shall start in Sec.~\ref{Sect.02} by covering the topic of  isospin impurities, which reflect the degree of isospin symmetry breaking. They are responsible for
isospin-forbidden rare decay modes and impact electroweak matrix elements
of decays, which probe fundamental symmetries.
Two examples of such decays will be discussed: the $0^+\rightarrow 0^+$ superallowed beta-decays
in Sec.~\ref{Sect.03} and  Fermi- and Gamow-Teller ground-state beta-decays in $T$=1/2 mirror nuclei in Sec.~\ref{Sect.04}.
Section~\ref{Sect.05} is devoted to  the mirror and triplet displacement energies sensitive to hadronic interactions, which break charge-symmetry and charge-independence. The phenomenon of proton-neutron pairing is covered in Sec.~\ref{Sect.06} in the context of a generalized DFT approach that is based on
proton-neutron mixed orbitals.
Finally, perspectives are given  in Sec.~\ref{Sect.07}.


\section{Isospin impurity}
\label{Sect.02}

The degree of isospin symmetry violation -- the isospin impurity -- is a result of a subtle balance between the attractive
short-range strong force and the repulsive long-range Coulomb interaction that polarizes the entire nucleus. Consequently,
its precise theoretical treatment requires the use of a {\it no-core\/} framework which, in heavier nuclei, is provided by
nuclear DFT.

The early attempts to evaluate the degree of isospin breaking, measured in terms of the isospin impurity $\alpha_C$, date back to the 1960s, see Ref.~\cite{[Aue83]} for a review.
These estimates, based on perturbation theory~\cite{[Sli65]} or analytically solvable hydrodynamical model~\cite{[Boh67]},
 were able to explain some gross features of $\alpha_C$ such as the steady increase of isospin mixing along the
$N=Z$ line with increasing $A$ or  quenching of  $\alpha_C$ with increasing $|N-Z|$.
These early models, however,  were not too quantitative.

\begin{figure}[thb]
\includegraphics[width=\columnwidth]{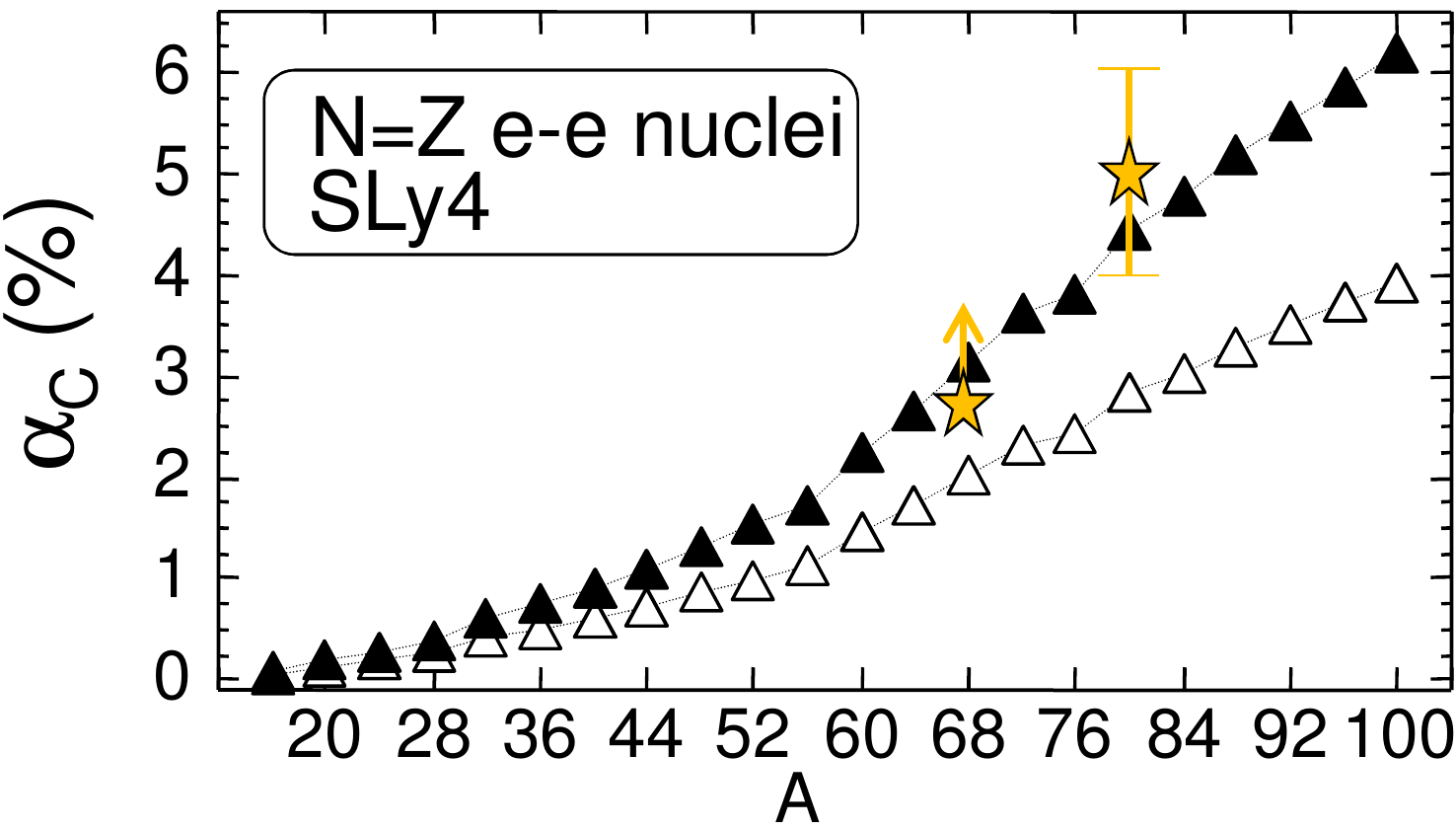}
\caption{\label{fig.1} Isospin impurities $\alpha_C$ in even-even $N=Z$ nuclei calculated
with SLy4  EDF~\protect\cite{[Cha97w]}. Full (open) triangles mark the AR (BR)
values, respectively. Note that, the mean-field estimates (BR)  are
quenched by $\approx$30\% with respect to the results of rediagonalization. Stars mark empirical values in
$^{64}$Ge~\protect\cite{[Far03]} (only the lower bound is known in this case) and $^{80}$Zr~\protect\cite{[Cor11x]}.
(Taken from Ref~\cite{[Sat13]}.)
}
\end{figure}

The accurate calculation of the isospin impurities
has been challenging.  This was early realized by Engelbrecht and Lermer~\cite{[Eng70]}
who pointed out that the self-consistent mean-field (MF) approaches cannot be directly applied because of the spurious mixing caused by the spontaneous isospin-symmetry-breaking effects.
To eliminate the  problem of spurious admixtures in the wave function, we have developed a no-core MR-DFT model involving
isospin projection, see also Refs.~\cite{[Cau80],[Cau82]}. The model employs self-consistent, isospin-broken
MF states $|\varphi \rangle$. Self-consistency is needed to ensure that the balance between the Coulomb force and the
strong interaction, represented in our model by the Skyrme energy density functional (EDF), are properly taken into account.
The MF state can be formally decomposed into good-isospin basis
$|T,T_z\rangle$:
\begin{equation}\label{mix}
|\varphi \rangle = \sum_{T\geq |T_z|}b_{T,T_z}|T,T_z\rangle,
\quad \sum_{T\geq |T_z|} |b_{T,T_z}|^2 = 1,
\end{equation}
where $T$ and $T_z=(N-Z)/2$ are the total isospin and its third component, respectively,
The mixing coefficients $b_{T,T_z}$ can be calculated using the states
\begin{equation}\label{Tbasis}
|T,T_z\rangle = \frac{1}{\sqrt{N_{TT_z}}}
\hat P^T_{T_z T_z}  |\varphi \rangle
\end{equation}
obtained by isospin projection after variation.

To assess the isospin mixing, one has to rediagonalize the total Hamiltonian ${\hat H}$ involving strong interaction plus the Coulomb term
in the space spanned by the good-isospin basis (\ref{Tbasis}):
\begin{equation}\label{mix2}
|n,T_z\rangle = \sum_{T\geq |T_z|}a^n_{T,T_z}|T,T_z\rangle ,
\end{equation}
where $n$ enumerates the  eigenstates $|n,T_z\rangle$ of ${\hat H}$. The value of  $n=1$, corresponds to the isospin-mixed  ground
state (g.s.).  The  g.s.\ isospin-mixing parameter obtained after rediagonalization (AR) is defined as:
\begin{equation}\label{alpha_C}
\alpha_C = 1- |a^{n=1}_{|T_z|,T_z}|^2.
\end{equation}
Comparison of $\alpha_C$ with  the quantity  $\alpha_C^{\textrm{(BR)}} = 1- |b_{|T_z|,T_z}|^2$ obtained before rediagonalization (BR)
provides direct information about the  spuriosity of MF solutions. As shown in Fig.~\ref{fig.1}, the isospin impurities
exceed  MF values by almost 30\%.

\begin{figure}[thb]
\begin{center}
\includegraphics[width=0.8\columnwidth]{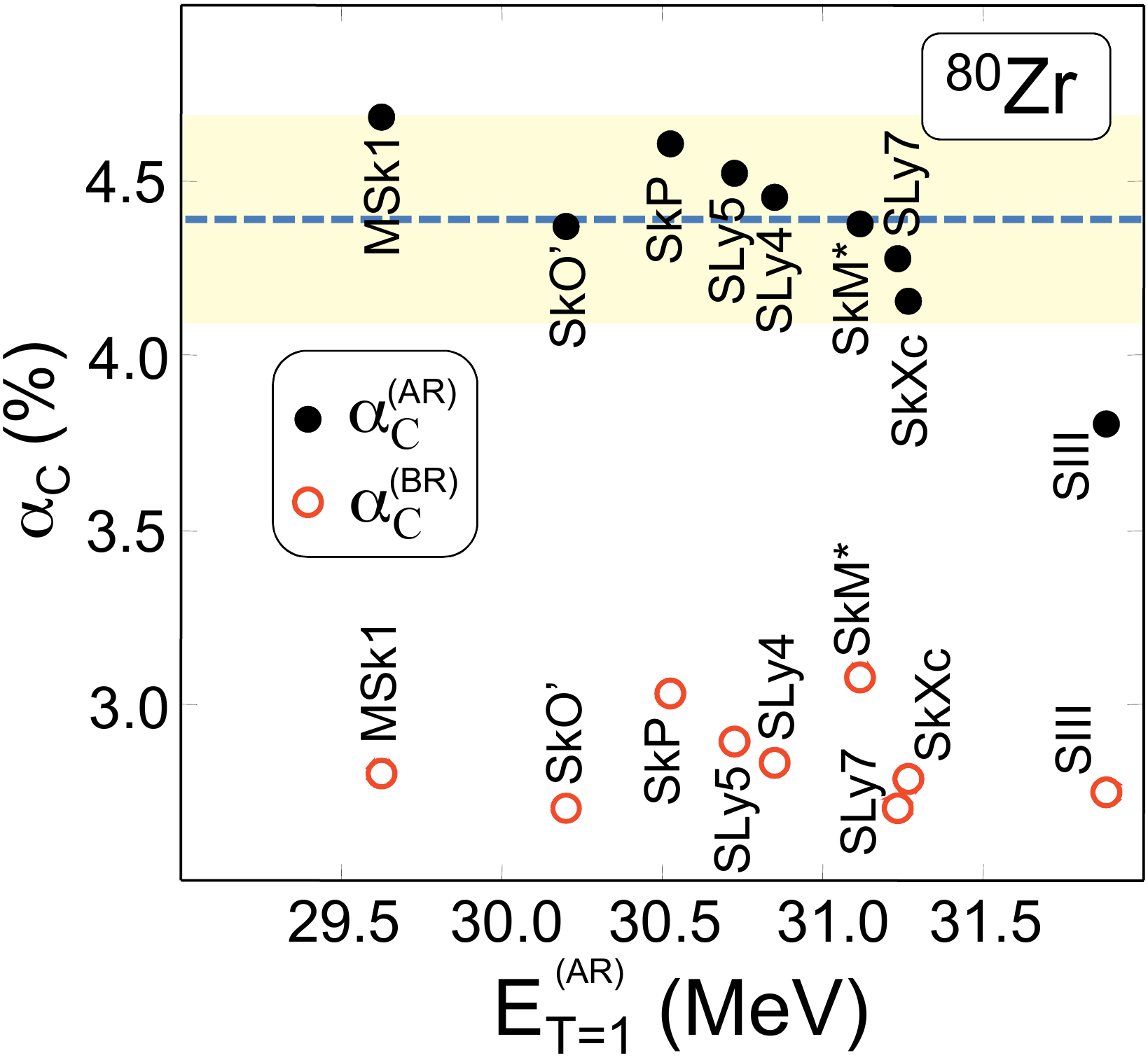}
\end{center}
\caption{\label{fig.2} Calculated isospin impurities in $^{80}$Zr versus excitation energy of the doorway state.
The results were obtained using the isospin projected DFT formalism with nine Skyrme EDFs. Filled (open) circles refer to AR (BR) results, respectively. The AR results lie within
experimental uncertainty limits of Ref.~\protect\cite{[Cor11x]} while the BR values are ruled out by the data. (Taken
from Ref.~\cite{[Sat11d]}.)
}
\end{figure}

The empirical information on isospin impurities in heavier systems is both scarce and uncertain.
The values of $\alpha_C$  extracted from a forbidden E1 transition in $^{64}$Ge~\cite{[Far03]} or
from the decay of giant dipole resonance  in $^{80}$Zr~\cite{[Cor11x]} are consistent with the predictions
of the isospin-projected DFT but the experimental uncertainties are too large to discriminate between different EDF
parameterizations. Indeed, as shown in Fig.~\ref{fig.2}, the spread in $\alpha_{\rm C}$ obtained with nine Skyrme EDFs,
$\bar{\alpha}_{{\rm C}}^{{\rm TH}} \approx 4.4\% \pm 0.3\%$, is relatively small and lies well within
the experimental uncertainty limits. The figure shows the calculated impurities versus
excitation energy $\Delta E_{{\rm T=1}}$ for the lowest $T=1$ state in even-even $N=Z$ nuclei, which is referred to as the {\it doorway state\/}. It is seen that the larger excitation energy of the  doorway state the smaller impurity, with exception of the SkO' result.

\begin{figure}[thb]
\begin{center}
\includegraphics[width=0.8\columnwidth]{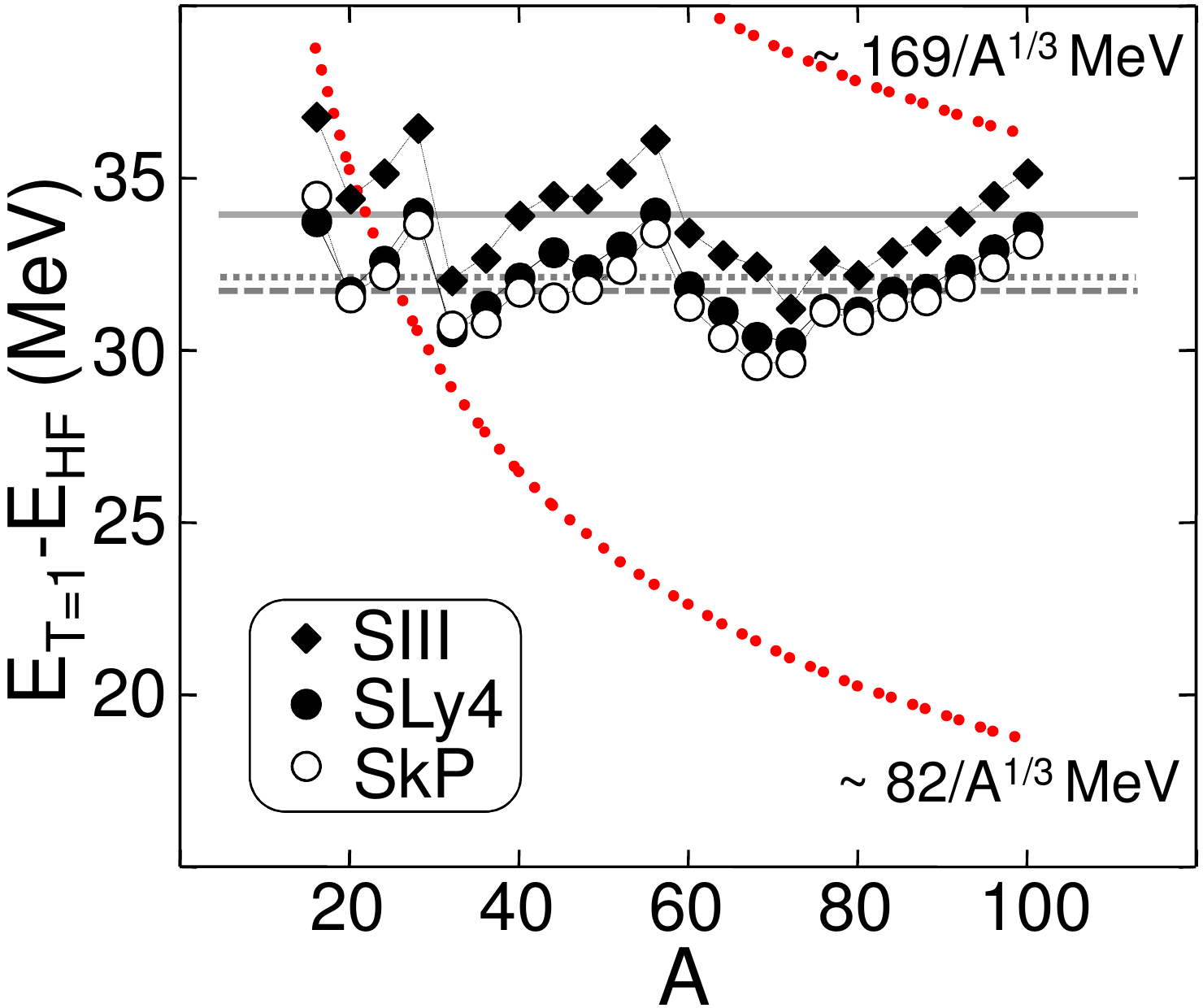}
\end{center}
\caption{\label{fig.3}
Excitation energies of the doorway states $E_{{\rm T=1}}$ in the even-even $N$=$Z$
nuclei relative to the g.s.\ energies $E_{\rm HF}$ obtained  with  SIII (diamonds),
SLy4 (dots), and SkP (circles) Skyrme EDFs. Horizontal lines mark the mean values.
The estimates based on the
perturbation theory~\cite{[Sli65]} and  the hydrodynamical model \cite{[Boh67]} are indicated by thick dotted lines.
(Adopted from Ref.~\cite{[Sat10]}.)
}
\end{figure}

Figure~\ref{fig.3} shows the calculated excitation
energy of the doorway state for the three  Skyrme EDFs: SIII~\cite{[Bei75]}, SLy4~\cite{[Cha97w]}, and SkP~\cite{[Dob84]}.
The self-consistent values of the excitation energies of the doorway state are compared  with the estimate $\Delta E_{{\rm T=1}}\approx 2\hbar\omega \approx 82/A^{1/3}$\,MeV of
Ref.~\cite{[Sli65]} based on the perturbation theory,
 and with the hydrodynamical estimate $\Delta E_{{\rm T=1}}\approx 169A^{1/3}$\,MeV of Ref.~\cite{[Boh67]}. It is seen that the simple estimates
of $\Delta E_{{\rm T=1}}$ vastly differ from the self-consistent results,  both in terms of the magnitude and $A$-dependence.
This indicates that the Coulomb mixing is a highly non-perturbative  effect that requires self-consistent treatment of the
interplay between short- and long-range interactions. That is why  back-of-the-envelope predictions of $\alpha_C$ are so unreliable.

\begin{figure}[thb]
\begin{center}
\includegraphics[width=0.8\columnwidth]{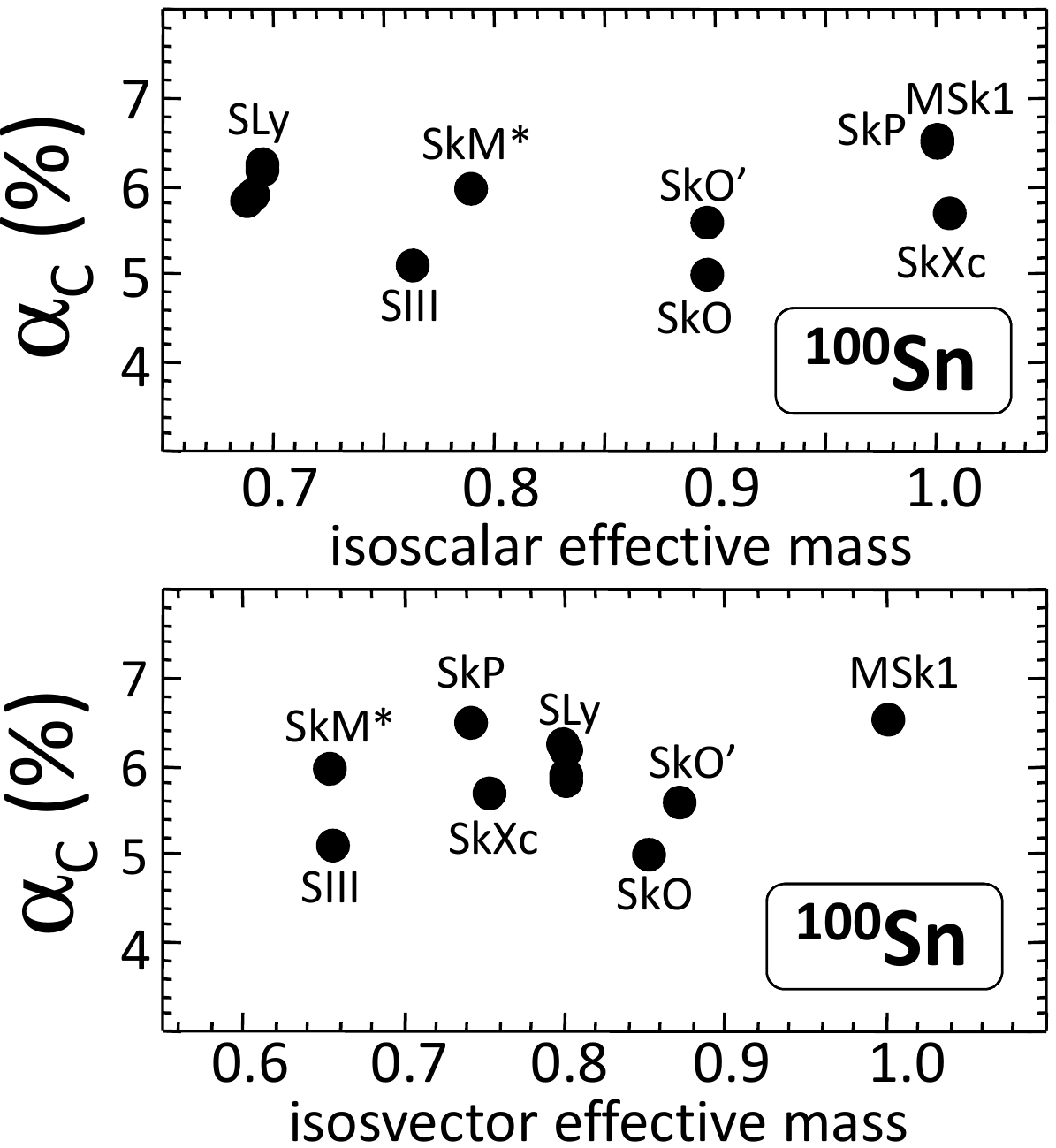}
\end{center}
\caption{\label{fig.4}
Isospin impurity $\alpha_{\rm C}^{\rm (AR)}$ in $^{100}$Sn calculated for several Skyrme EDFs plotted as a function of the isoscalar (top) and isovector (bottom) effective mass.
}
\end{figure}
Excitation energies of the doorway states and, in turn, isospin impurities, depend on  EDF parameterization.
It is, however, not at all obvious what EDF components   are responsible for the systematic differences seen in Fig.~\ref{fig.3}.
Indeed, attempts to correlate $\alpha_{\rm C}$ with various bulk properties of  Skyrme EDFs
turned out to be fairly inconclusive. For example, no correlation has been found with the
symmetry energy~\cite{[Sat11d]}, which is the primary
quantity characterizing the isovector properties of EDFs. As illustrated in Fig.~\ref{fig.4},
no clear correlation has been found between  $\alpha_C$ and the isovector and isoscalar effective mass.

\begin{figure}[htb]
\begin{center}
\includegraphics[width=0.8\columnwidth]{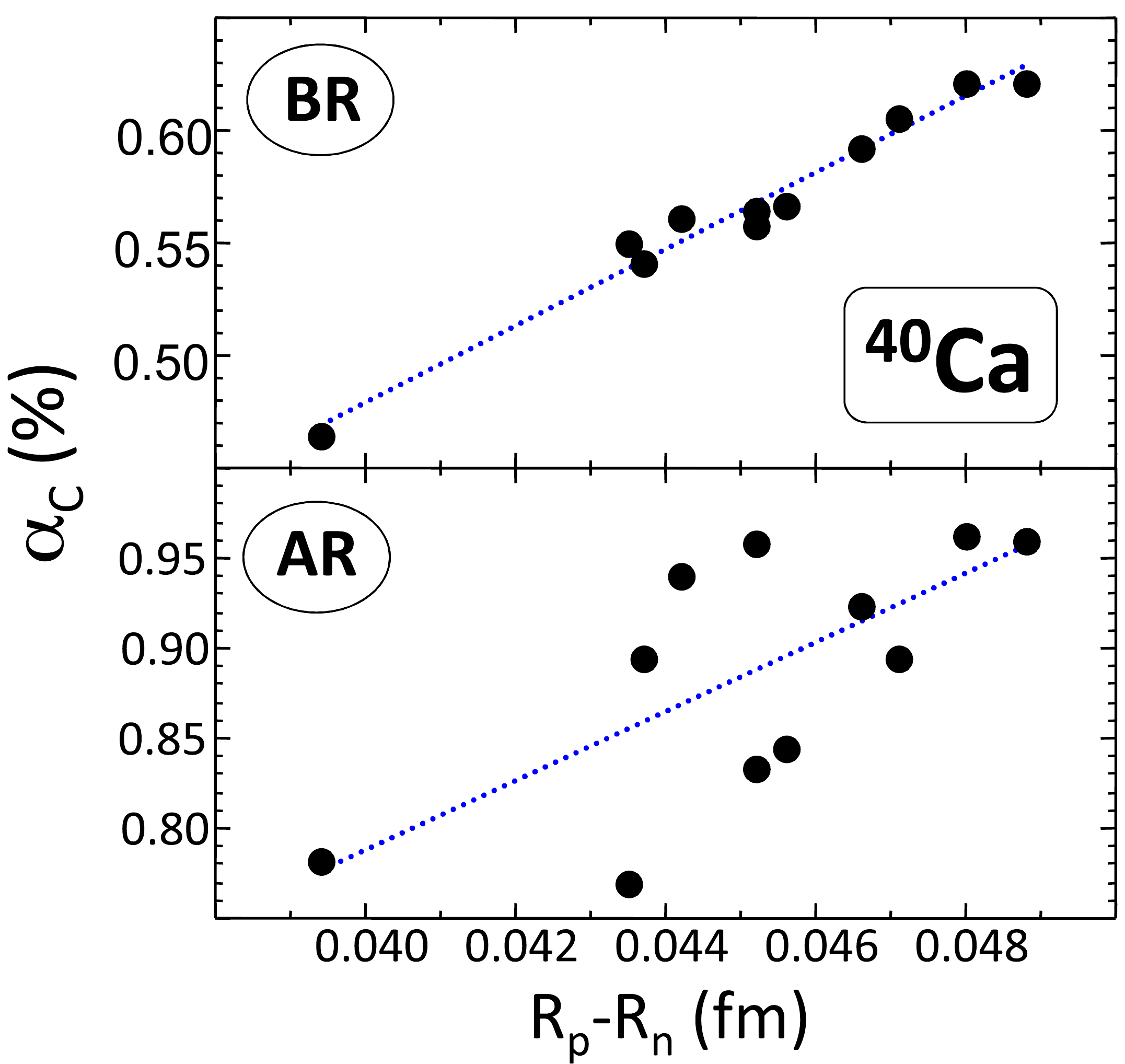}
\end{center}
\caption{\label{fig.5}
Coulomb impurities in $^{40}$Ca versus a difference of mean proton and neutron radii. Upper plot
shows mean-field values $\alpha_{\rm C}^{\rm (BR)}$. Lower plot shows true impurities $\alpha_{\rm C}^{\rm (AR)}$.
}
\end{figure}
We do find, however,  a strong correlation between  $\alpha_{\rm C}^{\rm (BR)}$  and the proton skin, i.e., the difference
between  proton and neutron  root mean square  radii, see Fig.~\ref{fig.5} and Ref.~\cite{[Sat09a]}.
The correlation deteriorates for $\alpha_{\rm C}^{\rm (AR)}$, indicating again that
the isospin mixing is a non-perturbative quantity; hence  difficult to estimate.

The formalism discussed above can be extended to incorporate  the angular-momentum projection.
In this case, the basis is created by applying the isospin ($\hat P^T_{T_z T_z}$) and angular-momentum ($\hat P^I_{MK}$) projection to  $|\varphi \rangle$:
\begin{equation}\label{ITbasis}
|I,M,K; T,T_z\rangle =   \frac{1}{\sqrt{N_{TT_z;IMK}}}
\hat P^T_{T_z T_z} \hat P^I_{MK} |\varphi \rangle ,
\end{equation}
where $M$ and $K$ denote the magnetic quantum numbers associated with
the laboratory and intrinsic $z$-axes, respectively \cite{[RS80]}.
Because $K$ is not conserved, the set (\ref{ITbasis}) is overcomplete. This problem
can be overcome by rediagonalization of the Hamiltonian in the so-called {\it collective space}, spanned for
each $I$ and $T$ by the {\it natural states\/}, $|IM;TT_z\rangle^{(i)}$~\cite{[Dob09d],[Zdu07a]}.
The  wave functions
\begin{equation}   \label{KTmix}
|n; IM; T_z\rangle =  \sum_{i,T\geq |T_z|}
   a^{(n)}_{iIT} | IM; TT_z\rangle^{(i)}
\end{equation}
obtained in the rediagonalization are labeled by the index $n$ and conserved quantum numbers $I$, $M$, and $T_z$.

\begin{figure}[htb]
\begin{center}
\includegraphics[width=0.8\columnwidth]{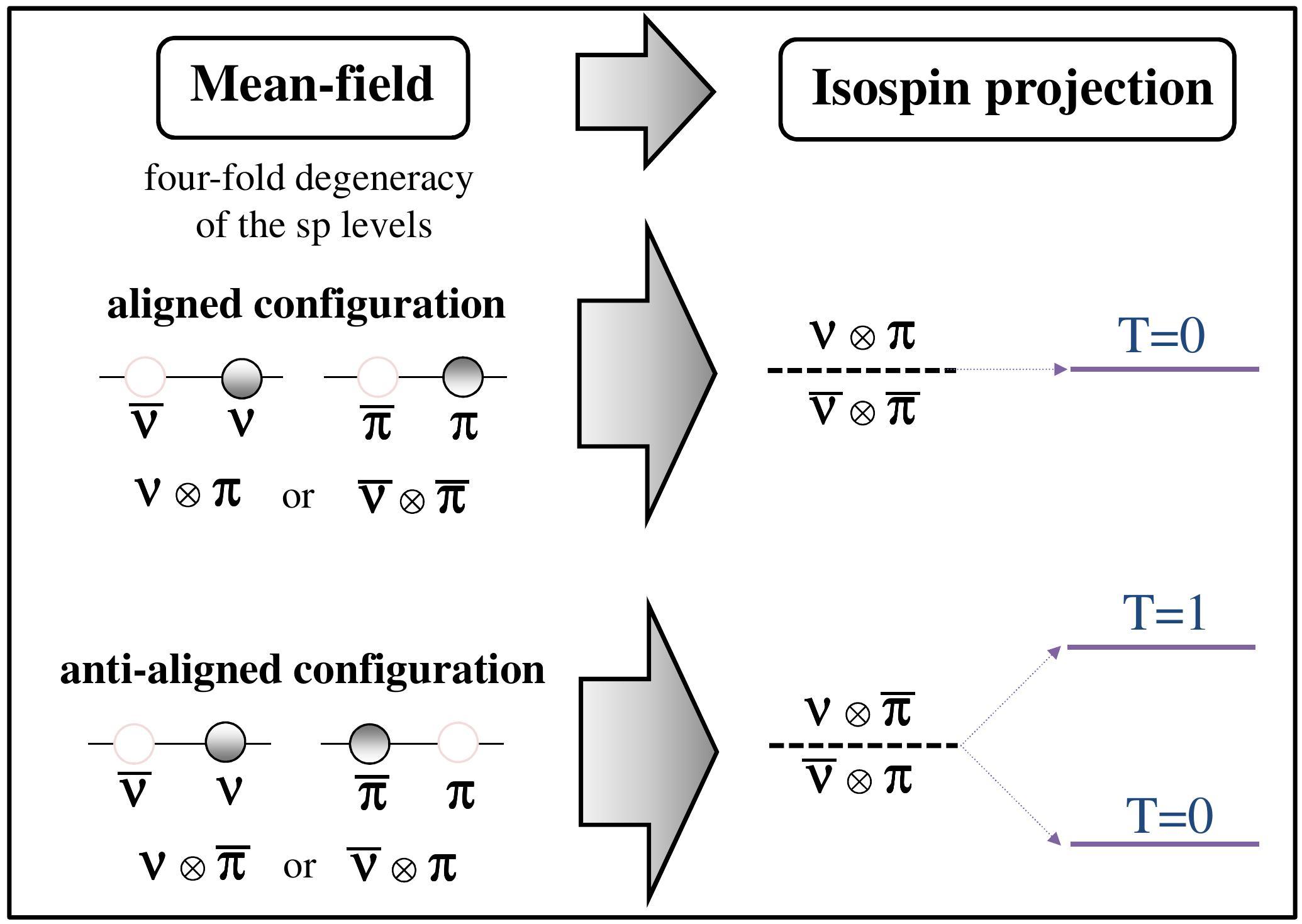}
\end{center}
\caption{\label{fig.6}
Left:  schematic  illustration of  two possible mean-field g.s.\ configurations in an odd-odd $N$=$Z$ nucleus.
Upper (lower) configuration is called aligned (anti-aligned). Right: the result of the isospin symmetry
restoration. The aligned configuration is isoscalar; hence, it is
insensitive to the isospin projection. The anti-aligned configuration represents a mixture of
$T$=0 and $T$=1 states. The  projection lifts the isospin degeneracy by lowering the $T$=0 level.
(Taken from Ref.~\cite{[Sat12]}.)
}
\end{figure}
\begin{figure}[htb]
\begin{center}
\includegraphics[width=0.8\columnwidth]{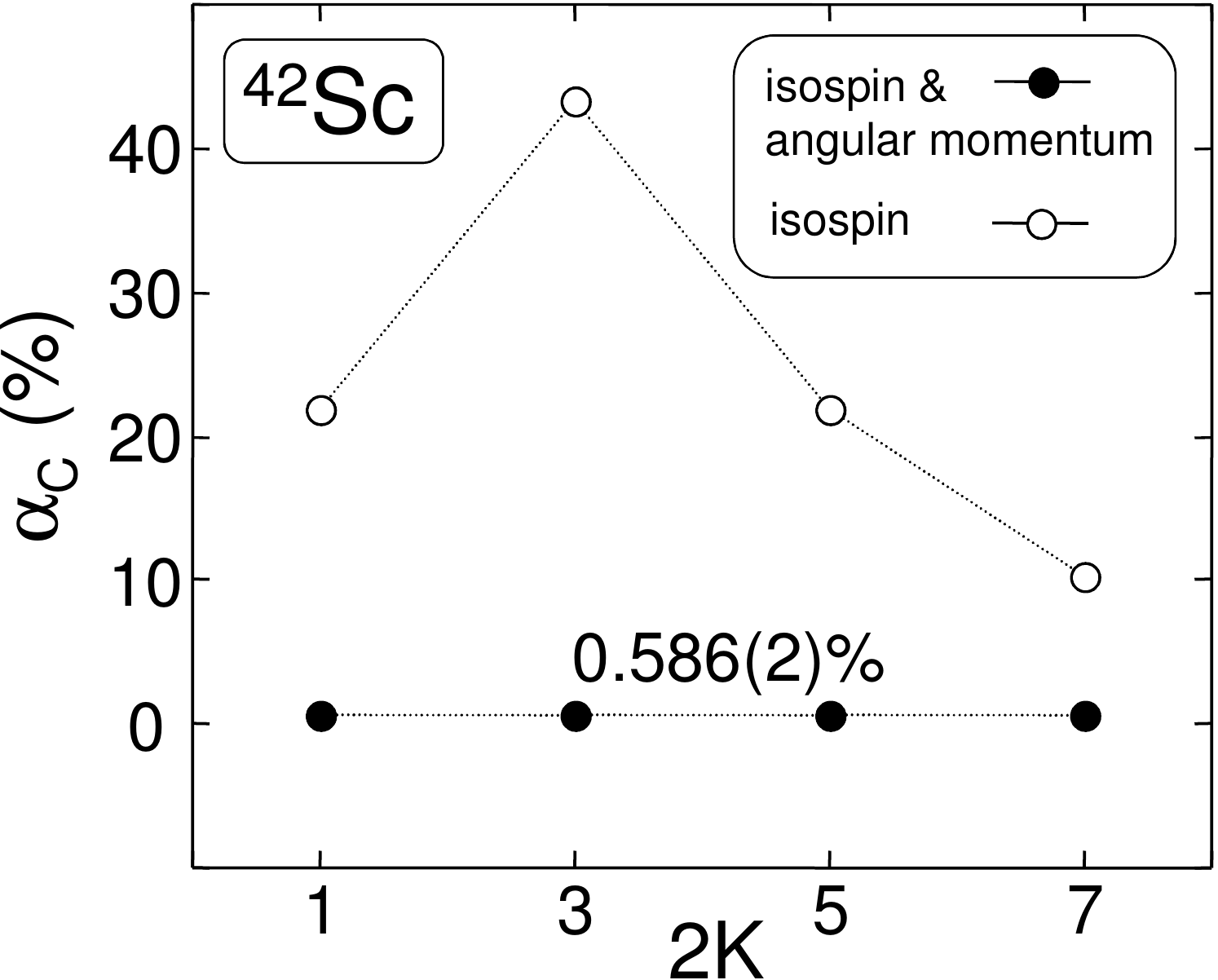}
\end{center}
\caption{\label{fig.7}
Isospin impurities in $^{42}$Sc, calculated for four antialigned
configurations $|\nu\bar{K} \otimes \pi K \rangle $  obtained by putting the valence neutron and proton in
opposite-$K$ Nilsson orbitals originating from the $f_{7/2}$ shell. Open and full dots show the results obtained in isospin-projected
and double-projected ($I=0, T=0$) variants, respectively. (Adopted from Ref.~\cite{[Sat11b]}.)
}
\end{figure}

The double-projected and isospin-projected approaches yield very similar g.s. isospin impurities  in even-even nuclei. However, there are configurations for which the model solely relying
on  isospin projection is insufficient. Among them there are the so-called
anti-aligned configurations in odd-odd $N=Z$ nuclei, which are crucial for
 calculations of ISB corrections to the superallowed beta-decay  involving the
$T=1, J=0^+$ states in odd-odd nuclei. These states are not representable by a single MF
configuration as shown schematically in Fig.~\ref{fig.6}. Indeed,
the odd proton and odd neutron can form either the aligned or anti-aligned configuration.
The aligned configuration represents an isoscalar $T=0, J\ne 0$ state. The anti-aligned configuration, however,  manifestly breaks the isospin symmetry being an equal mixture of  isoscalar
and isovector states. The calculations indicate~\cite{[Sat12]} that the $T$=0 and $T$=1 components projected
from the anti-aligned configuration strongly mix  through the Coulomb interaction
leading to unphysically large isospin impurities. Double-projected approach is free from such pathologies. This is demonstrated
in the lower part of Fig.~\ref{fig.7} for a representative case of $^{42}$Sc.

It can be shown~\cite{[Sat10]} that the isospin projection technique can be safely used
within nuclear DFT.
Augmenting the isospin projection with the angular-momentum projection leads to a number of theoretical issues. Not only  the numerical complexity of computations increases  but also the uncontrolled singularities appear
in the energy kernels ~\cite{[Sat11b]}; this essentially eliminates  a possibility to
use modern EDFs with density-dependent terms~\cite{[Ben03]}. To overcome this problem,
regularization schemes have been proposed~\cite{[Lac09],[Sat14b]} but they are difficult in practical implementations. Until a workable solution is developed, a practical option is
to use Hamiltonian-based EDFs, such as
the SV force~\cite{[Bei75]} augmented by the tensor terms, or the recently developed SLyMR
forces that include three- and four-body terms~\cite{[Sad13a]}.

All the existing  parameterizations of the density-independent Skyrme EDF are characterized by
a nonstandard   saturation mechanism driven by a very low isoscalar effective mass $\frac{m^*}{m}\approx 0.4$.
The low effective mass affects the overall performance of those forces, impairing such key properties as the symmetry energy,
level density, and level ordering. The low value of $\frac{m^*}{m}$ also impacts  the  isospin mixing.
In particular, in the case of $^{80}$Zr discussed above, the SV EDF yields
$\alpha_{\rm C} \approx 2.8$\%, which is considerably smaller than the mean value,
$\bar\alpha_{\rm C} \approx 4.4\pm 0.3$\%, obtained by averaging over the EDFs shown in Fig.~\ref{fig.2}.
The lack of a reasonable Hamiltonian-based Skyrme EDF is probably the most critical deficiency
of the current self-consistent approach. Nonetheless, one has to admit that the EDFs with low effective masses
perform surprisingly well when used in  the context of no-core-configuration-interaction (NCCI) extensions~\cite{[Bal14b],[Sat14],[Sat15],[Naz14]}


\section{Isospin-symmetry-breaking corrections to the superallowed $J^\pi=0^+, T=1  \rightarrow J^\pi=0^+, T=1$
beta decays}
\label{Sect.03}

The superallowed $J^\pi=0^+, T=1  \rightarrow J^\pi=0^+, T=1$ beta transitions are of particular importance for testing
various aspects of the electro-weak sector of the Standard Model. What makes these pure Fermi
decays so useful is the conserved vector current (CVC) hypothesis, which  states that the vector current is
not renormalized in the nuclear medium. This implies that the product of the statistical rate function $f$ and partial half-life $t$
for the superallowed $I=0^+,T=1 \rightarrow I=0^+,T=1$ Fermi beta-decay
should be nucleus-independent and equal to:
\begin{equation}\label{ft}
      ft = \frac{K}{G_{\rm V}^2 |M_{\rm F}^{(\pm )}|^2 } = {\rm const}\, ,
\end{equation}
where $K/(\hbar c)^6 = 2\pi^3 \hbar \ln 2 /(m_{\rm e} c^2)^5 =
8120.2787(11)\times 10^{-10}$\,GeV$^{-4}$s  is a universal constant; $G_{\rm V}$ stands for the vector
coupling constant for semi-leptonic weak interaction; and $M_{\rm F}^{(\pm )}$ is the nuclear matrix element of the isospin  operator $\hat T_{\pm}$.

In reality, the relation (\ref{ft}) holds only approximately and must be slightly amended
to account for radiative processes and isospin-symmetry breaking effects. Fortunately, the
radiative and ISB corrections are small, of the order of a few percent, and this allows us to
express the rate conveniently as:
\begin{equation}\label{ftnew}
   {\cal F}t \equiv ft(1+\delta_{\rm R}^\prime)(1+\delta_{\rm NS} -\delta_{\rm C})
  = \frac{K}{2 G_{\rm V}^2 (1 + \Delta^{\rm V}_{\rm R})},
\end{equation}
with the left-hand side being nucleus independent. In Eq.~(\ref{ftnew}),
$\Delta^{\rm V}_{\rm R} = 2.361(38)$\% stands for the
nucleus-independent part of the radiative correction~\cite{[Mar06b]}, $\delta_{\rm R}^\prime$ is
a transition-dependent ($Z$-dependent) but nuclear-structure-independent part of
the radiative correction~\cite{[Mar06b],[Tow08]}, and $\delta_{\rm NS}$ denotes
the nuclear-structure-dependent part of the radiative
correction~\cite{[Tow94],[Tow08]}. The ISB correction $\delta_{\rm C}$
is a many-body correction to the nuclear matrix element:
\begin{equation}\label{MFa}
|M_{\rm F}^{(\pm )}|^2 = 2 (1- \delta_{\rm C} ),
\end{equation}
accounting for the isospin-symmetry violation in the atomic nuclei, see Refs.~\cite{[Har09],[Tow10],[Tow10a],[Har14],[Har15]}
and references cited therein.

In spite of explicit dependence on theoretical input and the related uncertainties in the calculated radiative and
ISB corrections, the superallowed beta-decays provide today's most accurate value of the vector
coupling constant $G_{\rm V}$. Moreover, they enable a  very accurate verification
of the CVC hypothesis and provide  precise
 information on  the leading matrix element $V_{\rm ud} = G_{\rm V}/G_{\rm \mu}$ of the
CKM three-generation quark mixing matrix~\cite{[Cab63],[Kob73],[Tow10a],[Har14],[Oli14],[Har15]}. This is so because
the leptonic coupling constant, $G_{\mu}/(\hbar c)^3 = 1.16637(1)\times
10^{-5}$\, GeV${^{-2}}$, is  well known from the muon decay~\cite{[Oli14]}.
The uncertainty of $V_{\rm ud}$ extracted from the superallowed beta-decays is almost by an order of magnitude
smaller as compared to the values obtained from neutron or pion decays~\cite{[Oli14]}.

In search for  physics beyond the Standard Model, precision is of utmost importance.
Only those transitions that have $ft$-values measured with an accuracy $\le 0.3$\%  are
acceptable. The {\it canonical\/} set of transitions used over the last decade to test the Standard Model
consists of 13 such cases spreading over a wide range of masses from $A$=10 ($^{10}$Ca) to $A$=74 ($^{74}$Rb), see
~\cite{[Tow10a],[Har14],[Har15]}. With the CVC hypothesis  confirmed, it is possible to
extract  $V_{\rm ud}$ by averaging over several transitions. This feature makes the superallowed beta-decay strategy
very attractive. Recently, the  canonical set has
been extended by a measurement of  the beta-decay of $^{38}$Ca~\cite{[Par14],[Bla14]}. Moreover, the superallowed decay of
$^{18}$Ne is within experimental reach \cite{[Laf15]}.

The ISB corrections $\delta_{\rm C}$ were computed in diverse nuclear models, see Refs.~\cite{[Dam69b],[Har14],[Orm95a],[Sag96a],[Lia09],[Aue09],[Sat11a]}.
The standard in this field has been set by  Hardy and Towner (HT)~\cite{[Tow08],[Tow10],[Tow10a],[Har14],[Har15]}
who have used the nuclear shell model to account for the configuration mixing effect in conjunction with the mean-field potential  needed to account for a radial mismatch of proton
and neutron single-particle (s.p.) wave functions due to the long-range Coulomb polarization~\cite{[Orm95a]}. Such a description has certain practical advantages. For instance, it allows for an independent fine-tuning
of various model's ingredients. But it also leads to  internal inconsistencies. In particular, the HT model violates the SU(2) commutation rules for the
bare isospin operators as pointed out in Refs.~\cite{[Mil08],[Mil09]}. Unfortunately, the associated uncertainty is difficult to quantify.

The self-consistent  multi-reference DFT involving the isospin-
and angular-momentum projections~\cite{[Sat09a],[Sat10],[Sat11a],[Sat12]} is an interesting alternative to the fine-tuned phenomenological HT approach.
This is a {\it no-core\/} theory, which is capable of treating rigorously the rotational symmetry
and  explicit breaking of isospin symmetry. The correct  balance between the Coulomb and hadronic forces is maintained  by self-consistency
requirements. The approach allows for a
rigorous treatment of the Fermi matrix elements using bare isospin operators. The recently proposed NCCI extensions of the framework~\cite{[Sat14],[Sat15],[Sat15b]} allow to take into account more correlations due to inclusion of (multi)particle-(multi)hole  configurations.
As already mentioned, the weakest point of the formalism is the lack of a reasonable Hamiltonian-based Skyrme EDF.
One should stress, however, that $\delta_C$ does not depend on the absolute magnitude of isospin mixing but rather on  a difference between parent and daughter states \cite{[Aue09]}, i.e., it is a less sensitive quantity.

The MR-DFT calculations of $\delta_{\rm C}$ involving single reference states in parent and
daughter nuclei are carried out  as follows. The superallowed $0^+  \rightarrow 0^+$ Fermi beta-decay proceeds between
the g.s. $| I=0, T\approx 1, T_z = \pm 1 \rangle$ of the even-even nucleus  and its isospin-analogue partner $|I=0, T\approx 1, T_z = 0 \rangle$
in the $N=Z$ odd-odd nucleus. The corresponding transition matrix element is:
\begin{equation}\label{fermime}
M_{\rm F}^{(\pm )} =
\langle I=0, T\approx 1,
T_z = \pm 1 | \hat T_{\pm} | I=0, T\approx 1, T_z = 0 \rangle.
\end{equation}
Within the MR-DFT, the parent g.s.  is approximated  by a  projected state
\begin{equation}
  | I=0, T\approx 1, T_z = \pm 1 \rangle
  = \sum_{T\geq 1} c^{( \psi )}_{T} \hat P^T_{\pm 1, \pm 1}
     \hat P^{I=0}_{0,0} |\psi \rangle,
\end{equation}
where $|\psi \rangle$ is  the   g.s. of the even-even nucleus obtained in the self-consistent Hartree-Fock calculations.
The state $|\psi \rangle$ is unambiguously defined by filling in the pairwise
doubly-degenerate levels of protons and neutrons up to the Fermi level.
The daughter state  is approximated by
\begin{equation}\label{oddphi1}
  | I=0, T\approx 1, T_z = 0 \rangle
  = \sum_{T\geq 0} c^{( \varphi )}_{T} \hat P^T_{0, 0}
     \hat P^{I=0}_{0,0} |\varphi \rangle,
\end{equation}
where  the self-consistent Slater determinant $|\varphi \rangle \equiv |\bar \nu \otimes \pi \rangle$ (or  $| \nu \otimes
\bar \pi \rangle$)
represents the anti-aligned configuration, selected by placing the odd neutron and the odd proton in
the lowest available time-reversed (or signature-reversed) s.p.\ orbits. The isospin-projected state shown in Fig.~\ref{fig.6}, based on  the $|\varphi \rangle$
configuration that manifestly breaks the isospin symmetry, is indeed a preferred way to access the $|T\approx 1, I=0\rangle$  states in odd-odd $N=Z$ nuclei.

The anti-aligned configurations in odd-odd $N=Z$ nuclei are not uniquely defined.
In the signature-symmetry-restricted calculations, there are in general three anti-aligned
Slater determinants with the s.p.\ angular momenta (alignments)
of the valence protons and neutrons pointing along the $Ox$, $Oy$, or $Oz$ axes of the
intrinsic shape defined by means of the long ($Oz$), intermediate ($Ox$), and short
($Oy$) principal axes of the nuclear mass distribution, respectively. Thus far, in spite of persistent
efforts, no self-consistent tilted-axis g.s. solutions have been found.
Various properties of these linearly dependent solutions, hereafter referred to as  $|\varphi_{\rm X}\rangle$,
$|\varphi_{\rm Y}\rangle$, and $|\varphi_{\rm Z}\rangle$,  are discussed in detail in Ref.~\cite{[Sat12]} and will not be repeated here.

\begin{figure}[htb]
\begin{center}
\includegraphics[width=0.8\columnwidth]{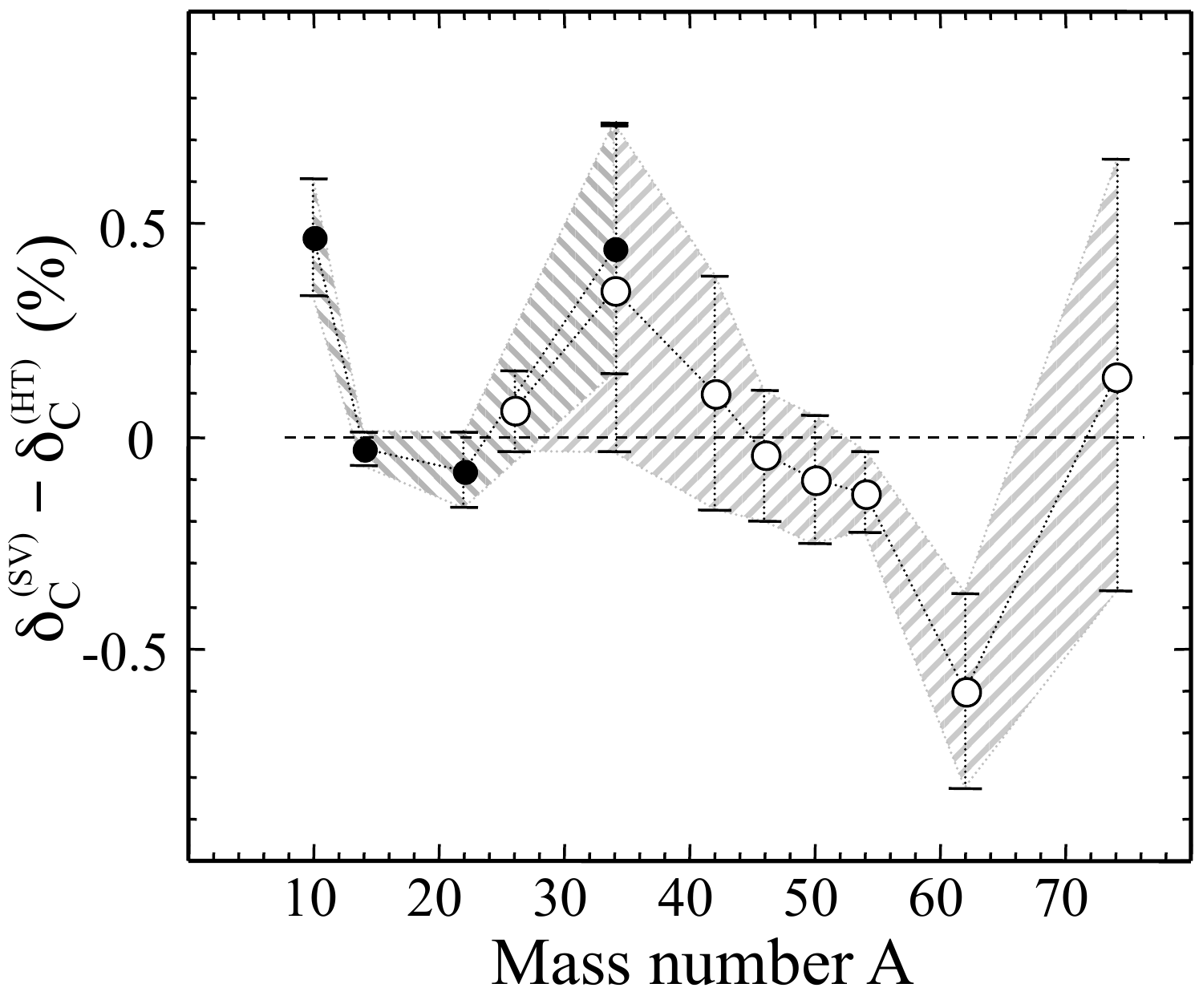}
\end{center}
\caption{\label{fig.8}
Differences between the ISB corrections to 12 precisely measured
superallowed $0^+\rightarrow 0^+$ decays (excluding $A$=38) calculated in Refs.~\cite{[Sat12]} (SV)
and~\cite{[Tow08]} (HT). Shaded band shows the combined SV+HT error. (Taken from Ref.~\cite{[Sat12]}.)
}
\end{figure}

In a MR-DFT double-projection approach based on a single reference state,
the only way to deal with the shape-current orientation ambiguity is to calculate three independent beta-decay matrix
elements corresponding to each orientation and average over the resulting values of $\delta_{\rm C}$.
Such an strategy was adopted in Ref.~\cite{[Sat12]}, which contains  the calculated corrections
for the canonical set of superallowed transition as well as  predictions for  yet-unmeasured cases. The calculated corrections are in reasonable agreement with the HT results as shown
in Fig.~\ref{fig.8}. Both sets of  corrections systematically overestimate the RPA results of
Ref.~\cite{[Lia09]}.

\begin{figure}[htb]
\begin{center}
\includegraphics[width=0.9\columnwidth]{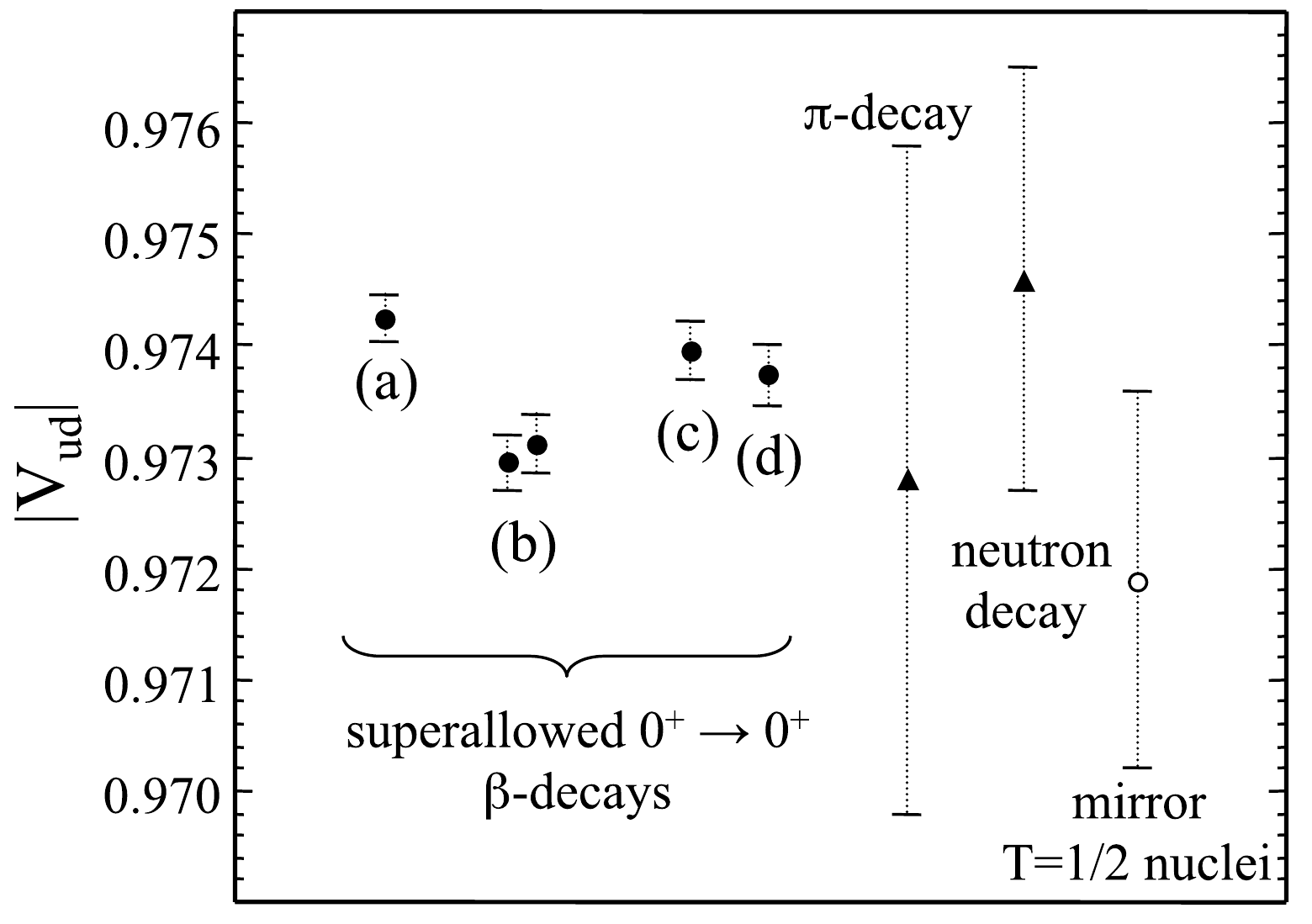}
\end{center}
\caption{\label{fig.9}
A survey of the $|V_{\rm ud}|$ values obtained using different methods: (a) Ref.~\cite{[Tow08]}; (b) Ref.~\cite{[Lia09]};  (c) Ref.~\cite{[Sat12]} with SV;  (d) Ref.~\cite{[Sat12]} with SHZ2. The values obtained form pion-decay~\cite{[Poc04]}, neutron-decay~\cite{[Nak10]}, and
 beta-decays in $T$=1/2 mirror nuclei~\cite{[Nav09a]} are shown for comparison. (Taken from Ref.~\cite{[Sat12]}.)
}
\end{figure}

With the set of ISB corrections obtained in Ref.~\cite{[Sat12]}, one obtains
$\overline{{\cal F}t} = 3073.6(12)$\,s, which gives $|V_{\rm ud}|= 0.97397(27)$.
This value is in excellent agreement with the HT result~\cite{[Tow08]},  $|V_{\rm ud}^{{\rm (HT)}}| = 0.97418(26)$,
and the central value obtained from the neutron decay,
$|V_{\rm ud}^{(\nu )}| = 0.9746(19)$~\cite{[Nak10]}. Taking the calculated $|V_{\rm ud}|$ and combining it with
$|V_{\rm us}| = 0.2252(9)$ and
$|V_{\rm ub}| = 0.00389(44)$  from the 2010 Particle Data
Group~\cite{[Nak10]}, one obtains
\begin{equation}\label{ckm}
       |V_{\rm ud}|^2 +  |V_{\rm us}|^2  + |V_{\rm ub}|^2 =  0.99935(67),
\end{equation}
which implies that the unitarity of the first row of the  CKM matrix  is satisfied with a precision better than 0.1\%.
A survey of the $|V_{\rm ud}|$ values obtained using different methods is shown in Fig.~\ref{fig.9}.

The solutions $|\varphi_{\rm X}\rangle$, $|\varphi_{\rm Y}\rangle$, and $|\varphi_{\rm Z}\rangle$,
differ by at most a few hundred keV in energy. Hence, there is no obvious preference for a choice of reference state. Moreover, the orientation-dependent effects originate, predominantly, from the time-odd fields of the nuclear MF. Hence, the orientation effects are present only in odd-odd nucleus adding up to a difference between parent and daughter nuclei. The averaging procedure applied in Ref.~\cite{[Sat12]} should be considered as a purely practical solution. The shape-current-orientation ambiguity has motivated us to extend the formalism to  allow for a dynamical mixing of states that are projected from low-lying (multi)particle-(multi)hole self-consistent MF configurations. In the context of the $\delta_{\rm C}$ calculations, the idea is to mix $0^+$ states projected from $|\varphi_{\rm X}\rangle$, $|\varphi_{\rm Y}\rangle$, and $|\varphi_{\rm Z}\rangle$ configurations, respectively. In such an approach,  the $I$=0 wave functions in an odd-odd $N$=$Z$ nucleus are approximated by:
\begin{equation}\label{oddphi}
  |n;\, I=0, T_z = 0  \rangle  =
  \sum_{i={\rm X,Y,Z}} \sum_{T'\geq 0} f^{(n;\,I=0,T_z=0)}_{i T'} \hat P^{T'}_{0, 0}
     \hat P^{I=0}_{0,0} |\varphi_i \rangle,
\end{equation}
where the  coefficients $f^{(n;\, I=0,T_z=0)}_{i T'}$
are obtained by solving the Hill-Wheeler-Griffin (HWG) equation, and $n$ enumerates the HWG eigenstates. The HWG equation has, typically,
three linearly independent solutions which, instead of $n$, can be conveniently labeled
by approximate isospin quantum number: $|I=0, T\approx 0, T_z = 0  \rangle,
|I=0, T\approx 1, T_z = 0  \rangle$, and $|I=0, T\approx 2, T_z = 0  \rangle$.

\begin{figure}[htb]
\begin{center}
\includegraphics[width=0.8\columnwidth]{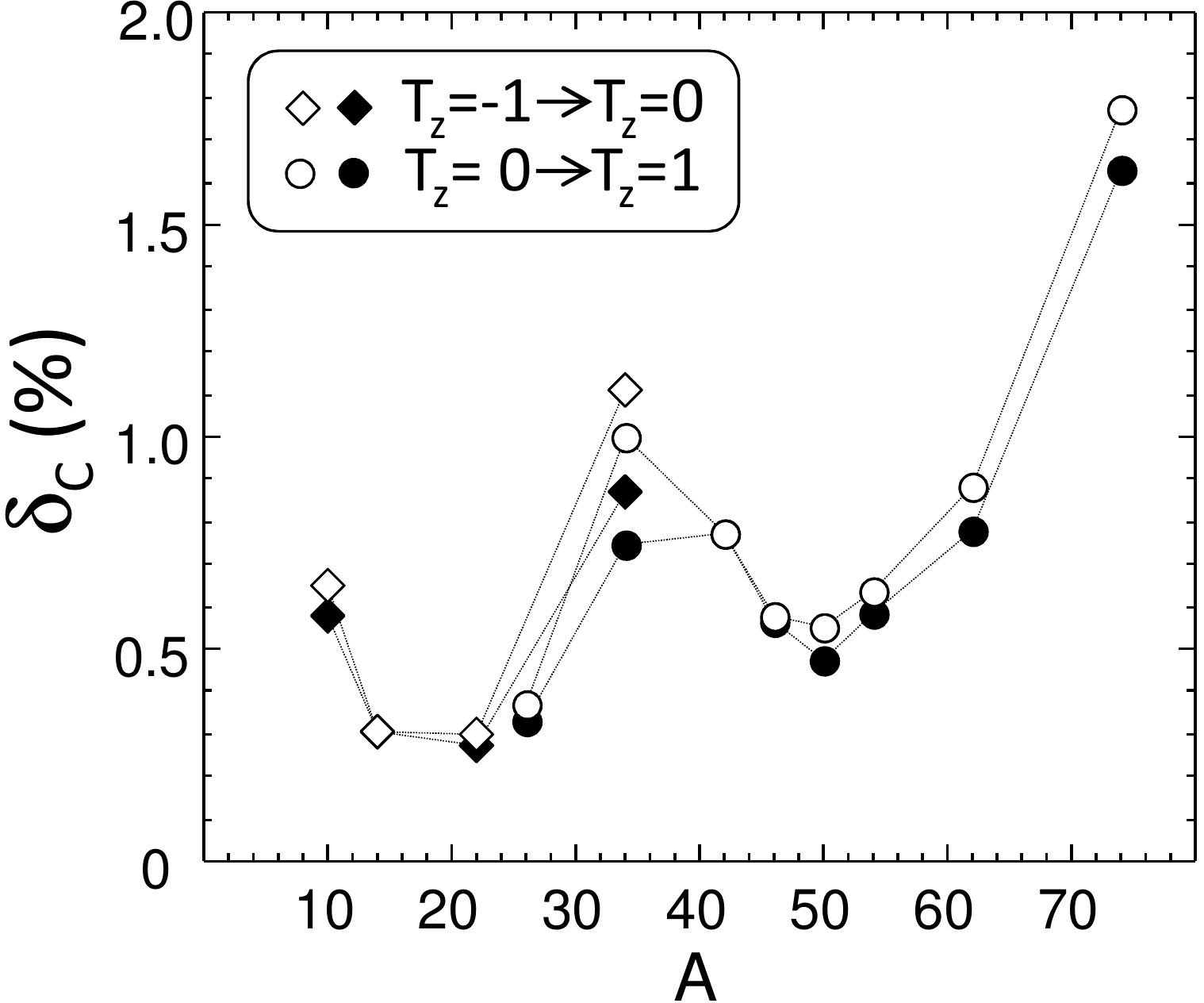}
\end{center}
\caption{\label{fig.10}
Comparison of the ISB corrections to the 12 canonical superallowed $0^+ \rightarrow 0^+$ transitions calculated in Ref.~\cite{[Sat12]} (open symbols) and in Ref.~\cite{[Sat15b]} (filled symbols). In the latter calculations, the $0^+$ states projected from
$|\varphi_{\rm X}\rangle$, $|\varphi_{\rm Y}\rangle$, and $|\varphi_{\rm Z}\rangle$ configurations
were mixed dynamically by solving the HWG equation. In the former,
matrix elements were calculated independently  for each orientation  and averaged afterwards.
}
\end{figure}
The new set of the ISB corrections calculated using this prescription is displayed in  Fig.~\ref{fig.10}; the detailed values can  be found in Ref.~\cite{[Sat15b]}. The improved corrections are somewhat smaller that the values of Ref.~\cite{[Sat12]} obtained by averaging over orientations. This difference, however,   has almost no influence on
$\overline{{\cal F}t}$ and $|V_{\rm ud}|$. Indeed the new values
$\overline{{\cal F}t} = 3073.7(11)$s and $|V_{\rm ud}|= 0.97396(25)$, calculated with the updated set of experimental
$ft$ values taken from Ref.~\cite{[Har14]}, almost perfectly match
the previous numbers \cite{[Sat12]}.

Towner and Hardy~\cite{[Tow10]} proposed a confidence level (CL) test to check a consistency of the calculated
ISB corrections. The underlying assumptions are: ({\it i\/}) validity of the CVC hypothesis up to at least $\pm$0.03\% and
({\it ii\/}) validity of the calculated nuclear-structure-dependent corrections $\delta_{\rm NS}$~\cite{[Tow94]}.
These two assumptions allow to calculate {\it empirical\/} corrections:
\begin{equation}\label{expdelt}
\delta_{\rm C}^{{\rm (exp)}}  = 1 + \delta_{\rm NS}
- \frac{\overline{{\cal F}t}}{ft(1+\delta_{\rm R}^\prime)}.
\end{equation}
By treating the value of $\overline{{\cal F}t}$ as an adjustable parameter, one can
bring $\delta_{\rm C}^{{\rm (exp)}}$ as close as possible to the calculated $\delta_{\rm C}$ by minimizing the appropriate $\chi^2$.
With the new set of corrections calculated in Ref.~\cite{[Sat15b]}, the
$\chi^2$ per degree of freedom ($n_d$=11) drops from $\chi^2 / n_d$=10.2~\cite{[Sat12]}
to 6.3. This number is still much larger than the values reported in Ref.~\cite{[Tow10]}, which are: 1.7 for the Damg{\aa}rd model
~\cite{[Dam69b]}, 0.4 for the shell-model with Woods-Saxon wave functions~\cite{[Har14]},
2.3 for the shell-model with Hartree-Fock wave functions~\cite{[Har14]}, and 2.1 for the relativistic DFT+RPA model of Ref.~\cite{[Lia09]}.
The main contribution to the $\chi^2$-value in our model can be associated
with the sudden increase
in $\delta_{\rm C}$ due to a single  $^{62}$Ga$\rightarrow$$^{62}$Zn transition, which
contributes more than 62\% to the total $\chi^2$ budget.


\section{Ground-state  beta-transitions in $T$=1/2 mirror nuclei}
\label{Sect.04}

The $T$=1/2 mirror nuclei offer an alternative way to test the CVC hypothesis~\cite{[Nav09a]}.
These nuclei decay via the mixed Fermi and Gamow-Teller (GT) transitions. Hence,
apart from the radiative and the ISB corrections,
the values of $G_{{\rm F}}$ and $V_{{\rm ud}}$ also depend  on the ratio of statistical rate functions $f_{{\rm A}}/f_{{\rm V}}$
for the axial-vector and vector interactions,  and the ratio $\rho \approx \lambda M_{{\rm GT}}/M_{{\rm F}}$ of nuclear matrix elements,
 where $\lambda =g_{{\rm A}}/g_{{\rm V}}$ denotes the ratio of axial-vector
and vector coupling constants.

The CVC hypothesis implies that the vector coupling constant is  $g_{{\rm V}} = 1$.
The axial-vector current is partially conserved; this implies  that the axial-vector coupling constant gets renormalized in the nuclear medium. The
effective axial-vector coupling constant, $g^{{\rm (eff)}}_A = q_s g_{{\rm A}}$, is quenched by an $A$-dependent factor $q_s$
with respect to the free neutron decay value $g_{{\rm A}} \approx -1.2701(25)$. Quenching factors deduced
from comparison of  large-scale nuclear shell model (NSM) calculations with experiment are: $q_s\approx 0.82$, 0.77~\cite{[Cho93]}, and 0.74~\cite{[Mar96]} in the $p$-, $sd$-, and $pf$-shell region, respectively. In the region $A\approx 130$, a  large  value of
$q_s \approx 0.57$ has been  extracted~\cite{[Cau12]}, see however~\cite{[Hor13]}. To account for the mass dependence of $q_s$ in shell-model calculations, a phenomenological expression
\begin{equation}
q_s=1-0.19 \left(\frac{A}{16}\right)^{0.35}
\end{equation}
has been proposed \cite{[Cho93]} for $A\le 40$.

The  origin
of the quenching is not fully understood. It is usually related to missing correlations in the wave function; truncation of model space; and -- in the context of {\it ab initio} models --
two-body currents~\cite{[Vai09],[Men11],[Eng14]}.
Much work in this area has been done in relation to the double-beta decay \cite{[Hol13],[Eng14],[Yao15]} and WIMP scattering \cite{[Klo13]}.
Recent study of the Ikeda sum rule in $\beta^-$ decays of $^{14}$C and $^{22,24}$O
performed in Ref.~\cite{[Eks14]} estimate quenching due to the two-body currents
to be $q_s=0.84-0.92$. This range is consistent with experimental data on $^{90}$Zr \cite{[Yak05],[Sas09]}.

Recently, a systematic study of both the GT and Fermi ground-state transitions in the $T$=1/2 mirror
$sd$- and $pf$-shell $T$=1/2 mirror nuclei with $17\le A \le 55$ have been carried out using the self-consistent NCCI approach~\cite{[Kon15]}. Within the NCCI model the wave function reads [cf.\ Eq.~(\ref{KTmix})]:
\begin{eqnarray}\label{NCCIwf}
|n; \, IM; \, T_z\rangle &=&  \sum_{i,j}
   a^{(n; IM; T_z)}_{ij} |\varphi_j;\, IM; T_z\rangle^{(i)}  \\
    &=&  \sum_{i,j} \sum_{K, T\geq |T_z|}
   f^{(n; IM; T_z)}_{ijKT} \hat P^T_{T_z T_z} \hat P^I_{MK} |\varphi_j \rangle \nonumber \, .
\end{eqnarray}
When compared to the isospin- and angular-momentum MR-DFT wave-function~(\ref{KTmix}), the NCCI wave function contains additional summation
over the configurations (Slater determinants) $|\varphi_j\rangle$. In the present implementation of the formalism, we use the same
Skyrme interaction to compute the configurations $|\varphi_j\rangle$ and to mix the isospin- and angular-momentum-projected states
 $|\varphi_j;\, IM; T_z\rangle^{(i)}$.

\begin{figure}[htb]
\begin{center}
\includegraphics[width=0.8\columnwidth]{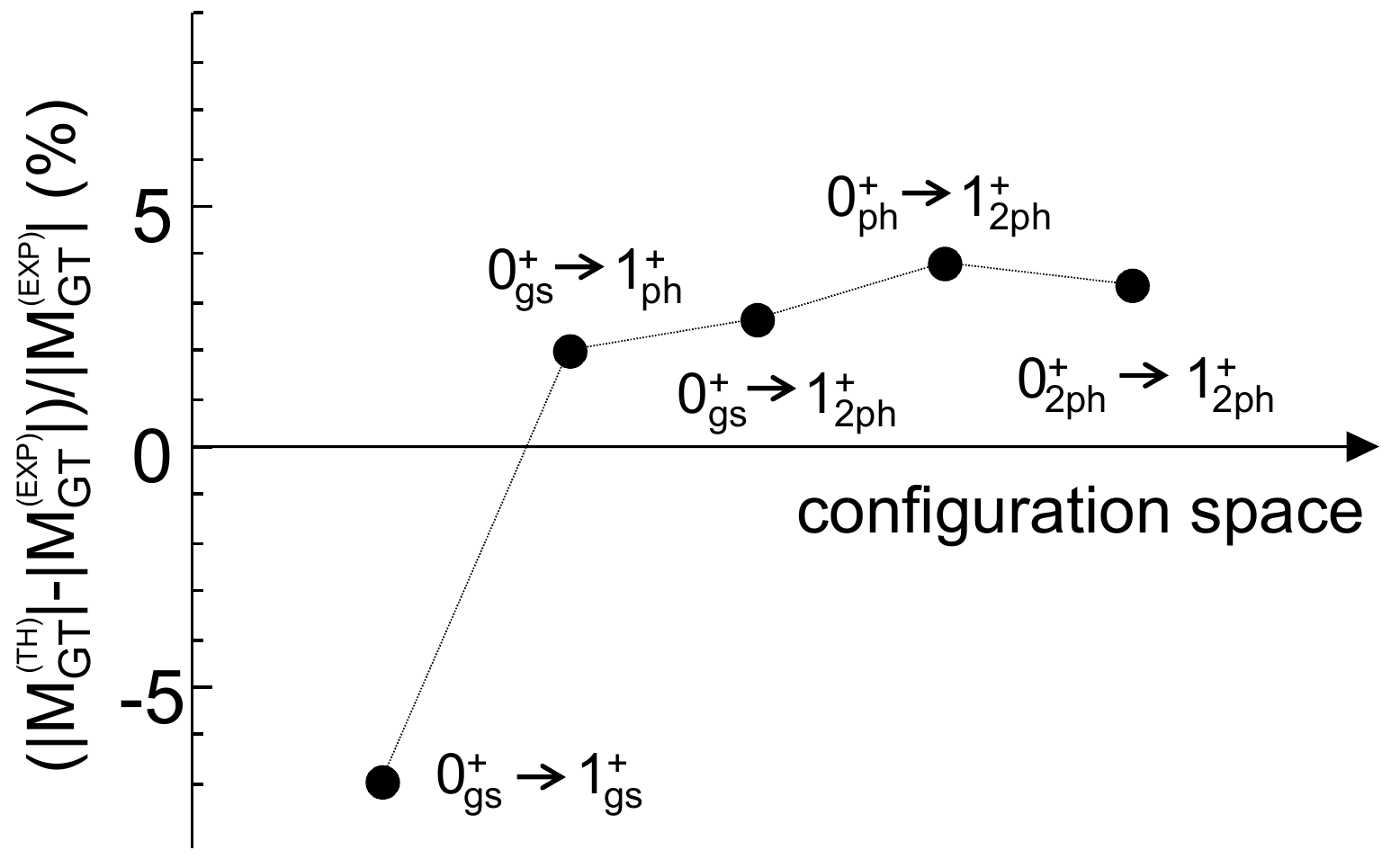}
\end{center}
\caption{\label{fig.11}
 GT $^6$He$\rightarrow$$^6$Li matrix element computed in NCCI relative to  experiment ~\cite{[Kne12]} as a
function of NCCI configuration space.
}
\end{figure}

The model was tested for the $^6$He$\rightarrow$$^6$Li GT beta decay. Matrix
element for this transition, $|M_{\rm GT}|$=2.1645(43), is precisely known from the recent measurement ~\cite{[Kne12]}.
Figure~\ref{fig.11} shows a difference between the calculated
and experimental GT matrix element as a function  of the NCCI configuration space considered.
The first point corresponds to the case of no configuration mixing.
Here, the wave functions $|0^+_{\rm gs}\rangle$ and $|1^+_{\rm gs}\rangle$ are projected
from the optimal (energy-wise) Slater determinants.
Next, keeping the parent wave function $|0^+_{\rm gs}\rangle$ fixed, we
enrich the configuration space of the daughter nucleus by admixing the $1^+$ states projected from the lowest particle-hole ($ph$) and the two lowest ($2ph$) particle-hole configurations:
$|1^+_{\rm gs}\rangle \rightarrow |1^+_{\rm ph}\rangle \rightarrow |1^+_{\rm 2ph}\rangle$.
This causes an increase of the matrix element  circa 3\% above the experimental value.
The extension of the configuration space  of $^6$He by admixing excited  $0^+$ states hardly impacts this result, see Fig.~\ref{fig.11}. This test indicates that NCCI is capable
of capturing the main features of the wave functions that are important for a reliable reproduction of the GT matrix element. Unfortunately, the model underbinds $^6$Li and overbinds $^6$He; hence we have not great confidence in its accuracy  for this specific GT transition. This is, however, not surprising as the current DFT approaches and their MR-DFT and NCCI extensions have not been optimized to experimental data in light nuclei.

\begin{figure}[htb]
\begin{center}
\includegraphics[width=0.7\columnwidth]{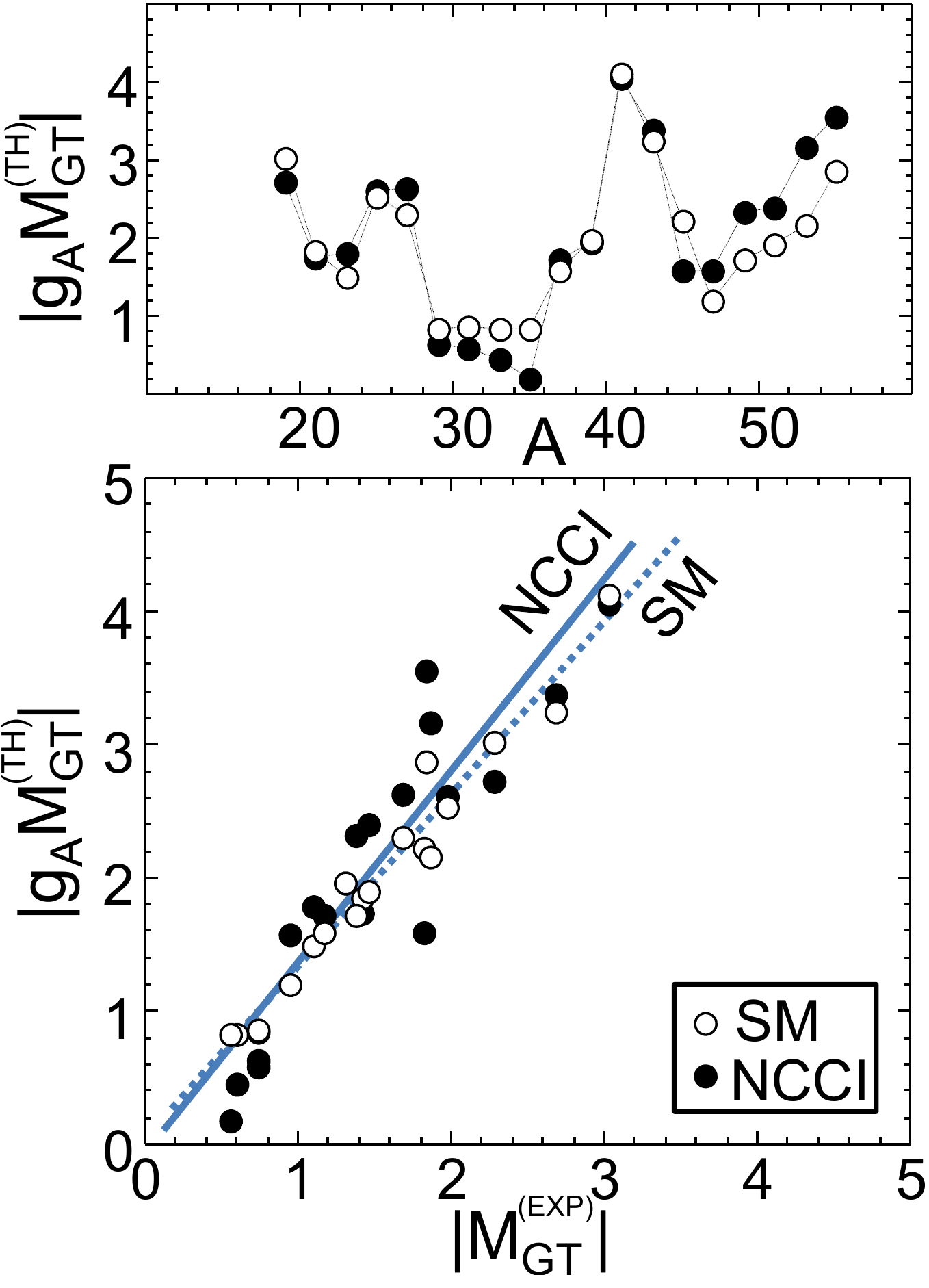}
\end{center}
\caption{\label{fig.12}
Top: ground-state GT matrix elements in $T$=1/2 medium-mass mirror nuclei calculated using  NCCI (filled circles)~\cite{[Kon15]} and shell-model (SM) approaches
(open circles)~\cite{[Bro85],[Mar96],[Sek87]}. Bottom:
a correlation between the theoretical and experimental matrix elements. Solid (dashed) line represents a  linear fit to the NCCI (SM) results.
}
\end{figure}

Theoretical ground-state GT matrix elements in the $T$=1/2 mirror $sd$- and $pf$-shell  nuclei ranging from $A$=17 till 55
are shown in Fig.~\ref{fig.12}. The NCCI results with the SV$_{\text{SO}}$ EDF were obtained using the HFODD solver~\cite{[Dob09d],[Sch12]} that has been augmented with
the NCCI module. The details of calculations follow Ref.~\cite{[Kon15]}.
It is seen that the NCCI results are fairly close to shell model predictions~\cite{[Bro85],[Mar96],[Sek87]}.
The consistency between these two theoretical approaches is visualized in
Fig.~\ref{fig.12}\,(bottom), which shows that both models correlate well
with experimental data. As discussed in Ref.~\cite{[Kon15]},
this similarity is quite surprising since the models differ in many aspects.

\begin{figure}[htb]
\begin{center}
\includegraphics[width=0.7\columnwidth]{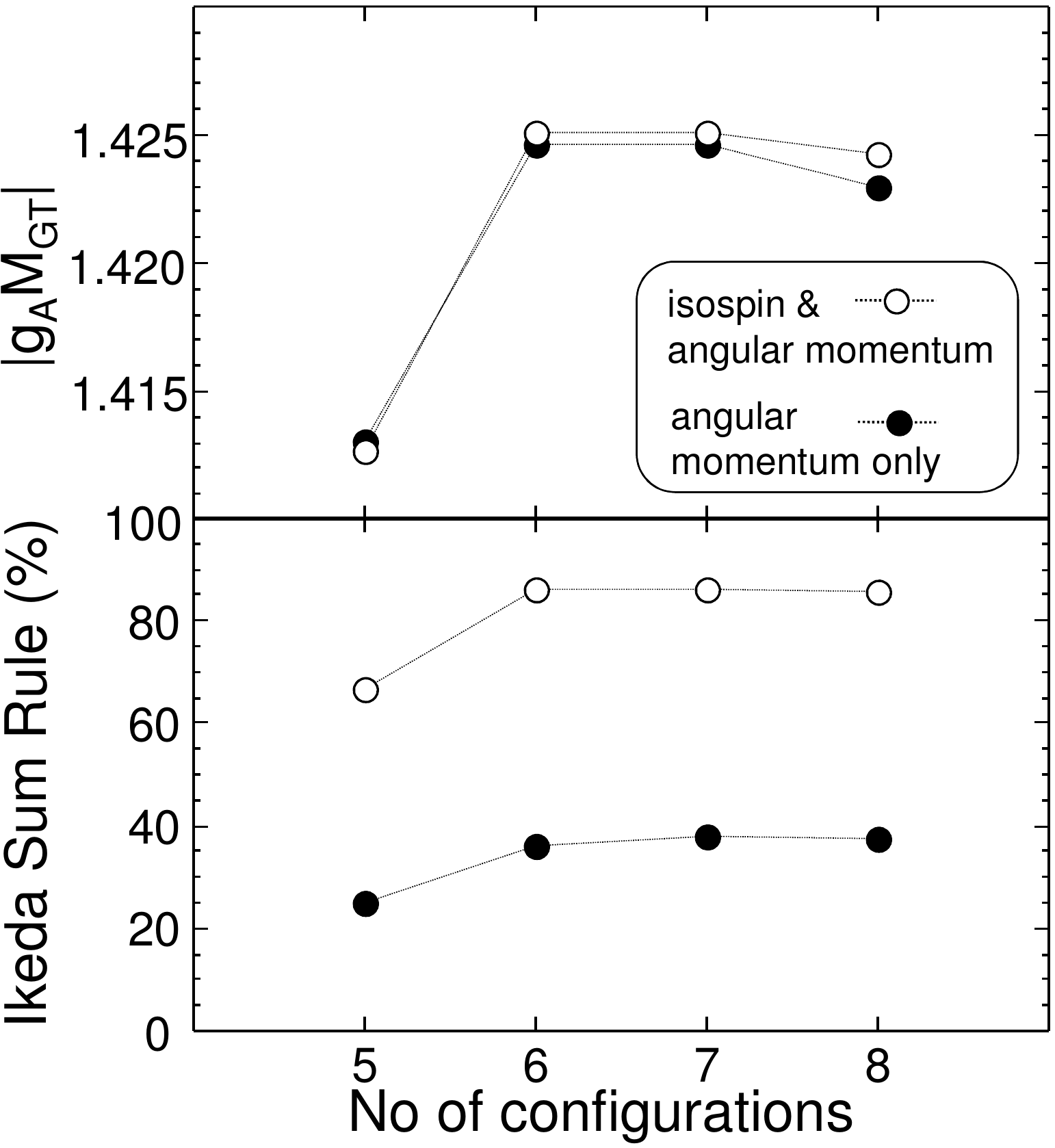}
\end{center}
\caption{\label{fig.13}
Top: GT matrix element for the g.s.$\rightarrow$g.s. transition $^{23}$Mg$\rightarrow$$^{23}$Na
versus the number of particle-hole configurations taken in the NCCI calculation
involving  angular-momentum projection (dots) and isospin- and angular-momentum projection (circles).
Bottom: the predicted Ikeda sum rule for the g.s. configuration of  $^{23}$Mg.
}
\end{figure}
The model-independent Ikeda sum rule \cite{[Ike64]} serves as an important indicator of the quality of theoretical models of GT decay.
In the  $A$=39, $T$=1/2 nuclei, which can be viewed as one-hole systems,
inclusion of all possible $ph$ excitations within the $sd$-shell space exhausts 99\% of the sum rule~\cite{[Kon15]}. Recently, we have performed similar calculations for the $A$=23, $T$=1/2
systems, which are complex, open-shell deformed nuclei. The calculations, involving eight deformed particle-hole
configurations, were performed using two variants of the NCCI model: the full model involving the isospin- and angular-momentum projection and its simplified variant involving only the angular-momentum projection.  The results are shown in Fig.~\ref{fig.13}.
As anticipated, both methods give almost identical matrix elements for the GS transition. This is because the effect of isospin mixing in the
ground states of  nuclei considered is very small. Note however, that the variant involving  the angular-momentum projection only captures mere
40\% of the Ikeda sum rule. At the same time, the full variant of the model accounts for the 85\% of the sum rule. The example shows
that the proper treatment of  isospin in excited states is critical for the sum rule evaluation.

\begin{figure}[htb]
\begin{center}
\includegraphics[width=0.8\columnwidth]{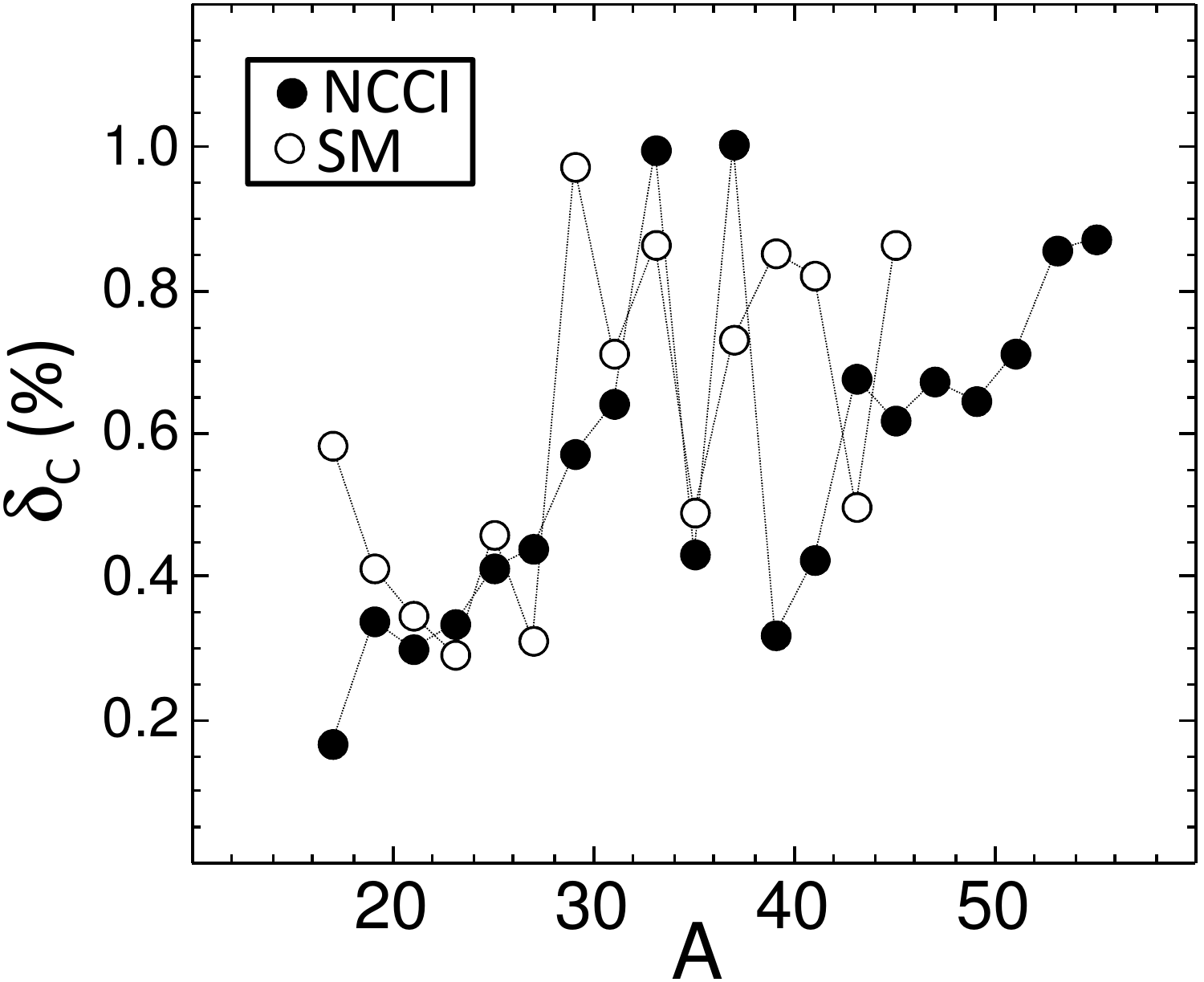}
\end{center}
\caption{\label{fig.14}
ISB corrections to the ground-state Fermi transitions in $T$=1/2 mirror nuclei obtained in NCCI~\cite{[Kon15]} (dots) and the SM+Woods-Saxon model of Ref.~\cite{[Sev08]} (circles).
}
\end{figure}
\begin{figure}[htb]
\begin{center}
\includegraphics[width=0.8\columnwidth]{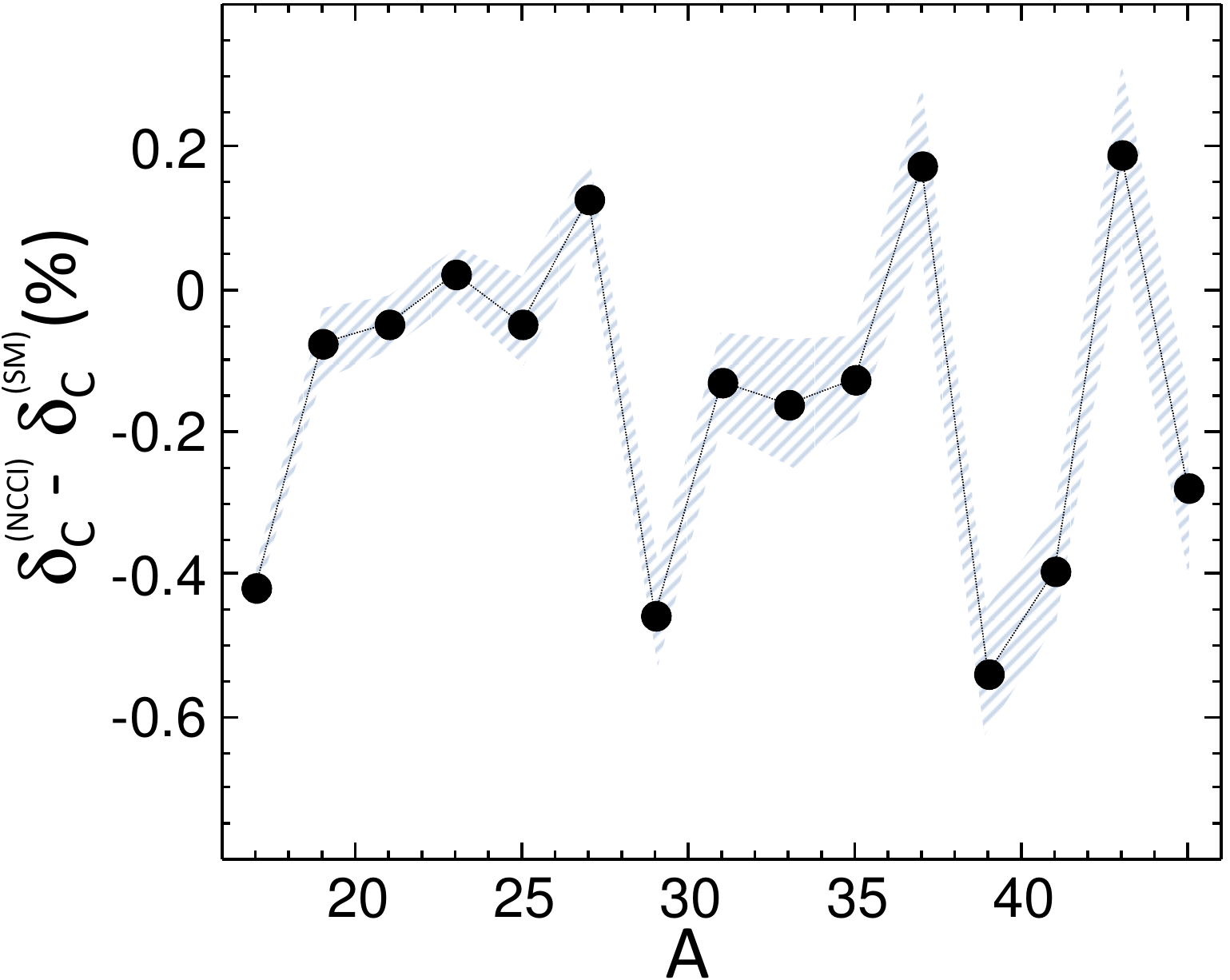}
\end{center}
\caption{\label{fig.15}
Difference between ISB corrections to the g.s. Fermi decays in $T$=1/2 mirror nuclei calculated with
 NCCI~\cite{[Kon15]} and the SM+Woods-Saxon model of Ref.~\cite{[Sev08]}.
The shaded area marks the theoretical uncertainty.
}
\end{figure}

The full variant of the model involving double-projection is necessary to compute the ISB corrections to the Fermi branch of
beta-decay in the $T$=1/2 mirror nuclei. These nuclei offer the independent way to verify the CVC hypothesis and to study the CKM matrix~\cite{[Nav09a],[Nav09b]}. A set of
the ISB corrections calculated with the MR-DFT technique was published in Ref.~\cite{[Sat11a]}, and
the NCCI results can be found in Table~1 of Ref.~\cite{[Kon15]}. The NCCI predictions are shown
in Fig.~\ref{fig.14} and compared to the SM results of Ref.~\cite{[Sev08]}.
It is  seen that the NCCI corrections,
although slightly smaller on average, are fairly consistent with  SM.
Figure~\ref{fig.15}
 shows a difference between these two sets of calculations. The shaded area marks the error band calculated
as $\sqrt{(\Delta\delta_{\rm C}^{{\rm (NCCI)}})^2+ (\Delta\delta_{\rm C}^{{\rm (SM)}})^2}$ under the assumption
of 10\% error on the NCCI results due  to a basis cut-off.

\section{Mirror and Triplet binding energy differences, and the Nolen-Schiffer anomaly} \label{Sect.05}

Strong interaction is, predominantly, invariant with respect to rotations in the isospin space.
This fact is rather well established experimentally and confirmed theoretically.
Indeed, most nuclear many-body approaches, including shell-model and  DFT
models, use interactions that are scalar in the isospace. The Skyrme interaction is a prime example of an isospin invariant
effective force that has long been  used  to compute various nuclear properties.

On the other hand, there  also exists a firm experimental evidence that the strong interaction contains small isospin-breaking components. For example, the nucleon-nucleon (NN) scattering data reveal small differences in phase shifts
and scattering lengths. In the $^1S_0$ channel, the scattering lengths for neutron-neutron ($nn$),  neutron-proton ($np$),
and proton-proton ($pp$) scattering are: $a_{nn}\approx -18.9$\,fm, $a_{np}\approx -23.7$\,fm, and
$a_{pp} \approx -17.3$\,fm, respectively \cite{[Mac01a]}.  A detailed analysis of scattering data, in particular phase shifts
and scattering lengths, shows that the \textit{nn} interaction is ~1\% stronger than \textit{pp} interaction and that the \textit{np} interaction is ~2.5\% stronger than the average of \textit{nn} and \textit{pp} interactions~\cite{[Mac01a]}.
(These figures include corrections for electromagnetic effects and refer solely to the strong nucleon-nucleon interaction.)
On the fundamental level, the isospin symmetry is broken due to different masses and electromagnetic interactions of $u$ and $d$ quarks \cite{[Bor15]}.

Albeit small, isospin violation considerably influences global properties of finite nuclei such as binding energies.  It is well known that models including isospin-invariant strong force and the Coulomb interaction alone  have difficulties in reproducing the  Mirror Displacement Energies (MDEs), the differences of binding energies of mirror nuclei (Nolen-Schiffer anomaly~\cite{[Nol69]}):
\begin{equation}
\mathrm{MDE}=BE\left(T,T_z=-T\right)-BE\left(T=,T_z=+T\right).\label{eq:MDE}
\end{equation}
These models has also problems in reproducing the Triplet Displacement Energies (TDEs). The TED indicator is defined as:
\begin{eqnarray}
\mathrm{TDE}=BE\left(T=1,T_z=-1\right)+BE\left(T=1,T_z=+1\right) \nonumber \\
            -2BE\left(T=1,T_z=0\right),\label{eq:TDE}
\end{eqnarray}
and is a measure of the binding-energy curvature within the isospin triplet~\cite{[Sat14],[Kan14],[Tu14],[Ben15]}.

It is customary to classify components of the nuclear force in terms of the SU(2) symmetry of isospace.  Here, one defines charge independence as invariance
under rotation in the isospin space. The charge symmetry, on the other hand, can be defined as an invariance under a rotation
by 180$^\circ$ about the $y$ isospin axis. Violation of this symmetry is referred to as charge symmetry breaking (CSB).

In~the language of NN force, CSB implies a
difference between $pp$ and $nn$ interactions in the same channel, $V_{nn} \ne V_{pp}$. Moreover, if in the isospin-triplet $T=1$ channel $V_{np} \ne (V_{nn} + V_{pp})/2$ the force is   charge-independence-breaking (CIB). The data on NN scattering lengths indicate that the strong NN  interaction must contain both the CSB and CIB strong components. On the fundamental level, the CSB mostly originates from the difference in masses of the proton and neutron which affects the kinetic energies and influences the boson exchange. The CIB originates mostly from the pion mass splitting, see Refs.~\cite{[Mac01a],[Mil95]}. In nuclear medium, the CSB and CIB components of the effective strong force
are accessible through the MDEs and TDEs, respectively.

A detailed classification of various components of the strong NN force was proposed by Henley and Miller~\cite{[Hen79],[Mil95]}.
They divided different components of the strong force into four classes.
According to this classification, the isospin invariant (i.e., isoscalar) NN interactions are called the class~I forces.
They commute with the total nuclear isospin operator, $\left[V_{{\rm I}}^{{\rm(NN)}},{\bf T}_A\right]=0$.  The class~II isotensor forces maintain charge symmetry, but break charge independence as $\left[V_{{\rm II}}^{{\rm (NN)}},{\bf T}_A\right]\neq0$. The class~III isovector forces break both the charge independence and charge symmetry, but are symmetric under interchange of nucleonic indices in the isospace. Such forces distinguish between the \textit{nn} and \textit{pp} systems and vanish in the \textit{np} system.  Forces of class~II and class~III do not mix isospin {\it only\/} at the two-body level; hence they
 contribute to the isospin mixing in finite nuclei.  Isovector forces of class~IV break both symmetries and mix isospin already at the two-body level. Those forces do not influence the \textit{nn} and \textit{pp} systems, but induce a spin-dependent isospin mixing effects in the \textit{np} system.

The isospin symmetry breaking components of  NN interactions were studied in
 \textit{ab inito} approaches~\cite{[Mac01a],[Wir13]}. Hadronic ISB terms of {\it effective}
NN interactions were investigated in the context of
 nuclear shell model~\cite{[Orm89a],[Zuk02],[Ben07a],[Qi08],[Kan12a],[Kan13],[Kan14],[Lam13]} and
 mean-field approaches~\cite{[Sag95a],[Sag96a],[Bro98a],[Bro00b]}. A note of caution is in order here. First,  any attempt to extract effective NN interactions from spectroscopic data should first account for the coupling to the many-body continuum~\cite{[Tho52],[Ehr51]} in the presence of isospin-conserving nuclear forces. If neglected, or not treated carefully, the continuum effects can alter the results of such analyses~\cite{[Gri02],[Mic10],[Yua14]}. Second, in-medium nuclear effective interactions (G-matrix) contain contributions from the Coulomb force. If the Coulombic contributions are not treated precisely during the renormalization procedure, they can result in the presence of CSB and CIB components, which can then be incorrectly labelled as ``hadronic".

To investigate the effect of hadronic CSB and CIB terms and their possible influence on
isospin mixing and ISB corrections, one can apply the local  Skyrme-DFT approach. To this end, one needs to generalize the formalism to the case of $pn$-mixed quasiparticle  states~\cite{[Per04]}.
Such an extension leads to isospin-invariant   EDFs that
depend explicitly on local  $pn$ densities and currents. Recently, this formalism was applied to the Hartree-Fock case  by admitting   $pn$ mixing in the particle-hole channel~\cite{[Sat13c],[She14]}.
Within this framework, the explicit ISB comes entirely from the electrostatic interaction.

In the next step, the $pn$-mixed  formalism can be extended to accommodate  the CSB and CIB hadronic components.
The $pn$-mixing is a necessary prerequisite that allows to study  these terms
in a fully self-consistent  manner. Since the discrepancies between the experimental and theoretical
MDEs and TDEs are small~\cite{[Bro00b],[Sat14]} they can be modelled in terms of  class~II (CIB) and class~III (CSB)  zero-range corrections to the conventional Skyrme force:
\begin{equation}
\hat{V}_{{\rm II}}(i,j)  =
\frac12 t_0^{{\rm II}}\, \delta\left({\bf r}_{ij}\right)
\left( 3\hat{\tau}_3(i)\hat{\tau}_3(j)-\vec{\tau}(i)\circ\vec{\tau}(j)\right),
\label{eq:class2}
\end{equation}
\begin{equation}
\hat{V}_{{\rm III}}(i,j)  =
\frac12 t_0^{{\rm III}}\, \delta\left({\bf r}_{ij}\right)
\left(\hat{\tau}_3(i)+\hat{\tau}_3(j)\right),
\label{eq:class3}
\end{equation}
where $t_{0}^{{\rm II}}$ and $t_{0}^{{\rm III}}$
are the ISB force parameters that are adjusted to reproduce empirical TDEs and MDEs.
The spin-exchange has been omitted as it leads to a trivial rescaling
of the parameters. The associated ISB contributions to the EDF are:
\begin{eqnarray}
\mathcal{H}_{{\rm II}} & = &
\frac{1}{2}t_0^{{\rm II}}
 \left(
\rho_n^2+\rho_p^2-2\rho_n\rho_p-2\rho_{np}\rho_{pn} \right. \nonumber \\
&\,& \left. -
\gras{s}_{n}^2-\gras{s}_{p}^2+2\gras{s}_{n}\cdot\gras{s}_{p}+2\gras{s}_{np}\cdot\gras{s}_{pn}
\right), \\
\mathcal{H}_{{\rm III}} & = &
\frac{1}{2}t_0^{{\rm III}}
\left(\rho_n^2-\rho_p^2 - \gras{s}_{n}^2+\gras{s}_{p}^2\right).
\end{eqnarray}

Note, that class II forces depend on the $pn$-mixed particle $\rho_{np}$ and spin $\gras{s}_{np}$ densities, respectively.
Hence, these forces can be included only within the $pn$-mixed DFT formalism. The class~III forces, on the other hand,
depend only on the standard $pp$ and $nn$ densities and can be, therefore, treated within the conventional  Skyrme-DFT approach.

In order to control the total isospin of the nucleus in the $pn$-mixing calculations, we use a three-dimensional isocranking approach \cite{[Glo04w]}. This technique is analogous to the well known cranking method in real space, which is
commonly used in high-spin physics. It is realized by adding the isocranking term to the
mean-field Hamiltonian $\hat {h}$,
\begin{equation}
\hat{h}'
=\hat {h} -\vec{\lambda} \circ \vec{t}
= \hat {h}  - \lambda_1\hat{t}_1-\lambda_2\hat{t}_2-\lambda_3\hat{t}_3,
\label{eq:isocranking}
\end{equation}
where $\vec{t}=\frac{1}{2}\vec{\tau}$ stands the s.p. isospin operator and $\vec{\lambda}$ is the isocranking frequency.
By adjusting the  frequencies $\vec{\lambda}$, one can control both
the length and direction of the isospin vector. In practical applications, $\vec{\lambda}$ is parameterized as follows:
\begin{equation}
\vec \lambda  
              =(\lambda^{\prime}\sin \theta^\prime \cos \phi,
              \lambda^{\prime}\sin \theta^\prime \sin \phi,
              \lambda^{\prime}\cos \theta^\prime +\lambda_{\rm off}). \label{eq:isofrequency}
\end{equation}
The offset $\lambda_{\rm off}\ne 0$ is introduced to facilitate
calculations with the Coulomb interaction~\cite{[Sat13c],[She14]}. By choosing the offset properly,
one can compensate for the effect of the electrostatic
interaction on the third component of the isocranking term. In this way, $\lambda_{\rm off}$ helps
to avoid s.p. level crossings while tilting the isocranking
axis and, consequently, keeps the expectation value of the total isospin fixed~\cite{[Sat01]}.

\begin{figure}[htb]
\begin{center}
\includegraphics[width=0.8\columnwidth]{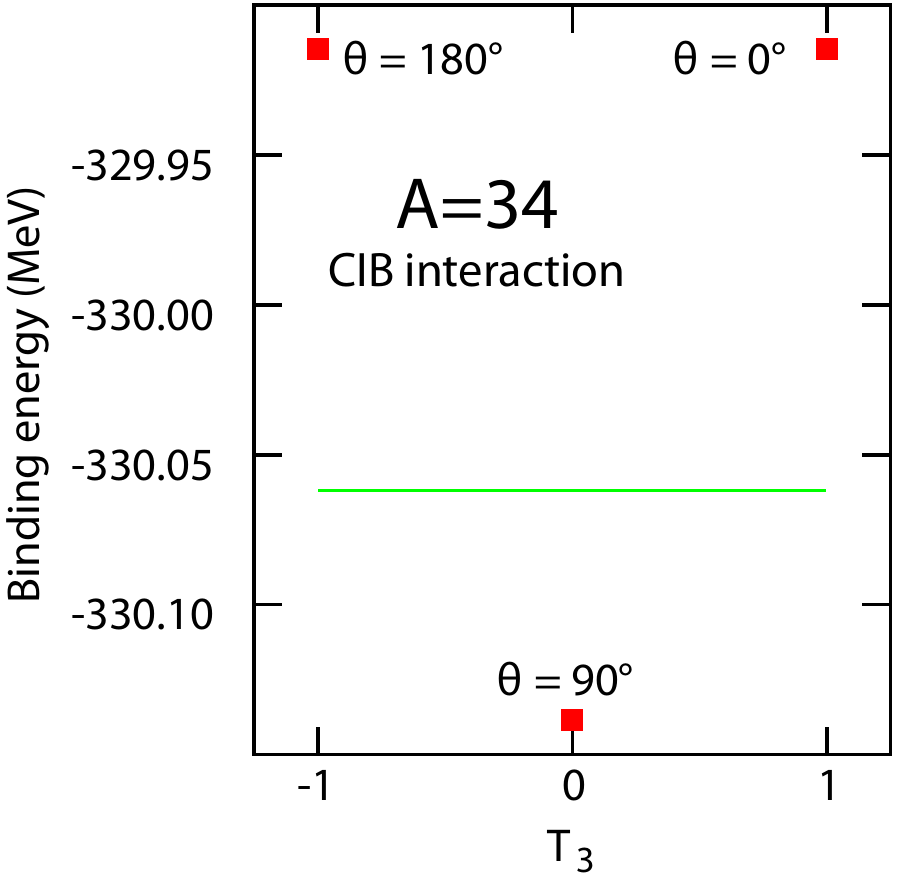}
\end{center}
\caption{\label{fig.16}
Binding energy for the $A$=34 isospin triplet calculated using the isocranking approach
with the isoscalar SV Skyrme force without Coulomb, augmented by the CIB (class II) interaction
(\ref{eq:class2}) with $t_0^{{\rm II}}$=20\,MeV.
}
\end{figure}

\begin{figure}[htb]
\begin{center}
\includegraphics[width=0.8\columnwidth]{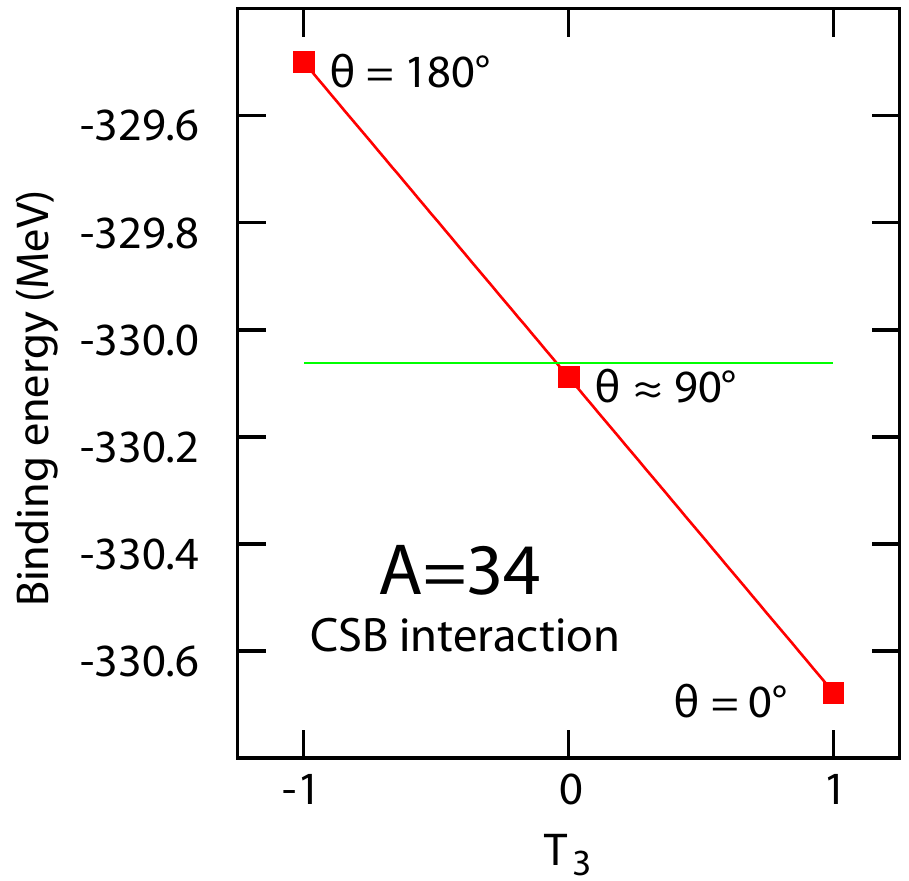}
\end{center}
\caption{\label{fig.17} Similar as in  Fig.~\ref{fig.16} but for the CSB (class III) interaction (\ref{eq:class3}) with $t_0^{{\rm III}} =-8$\,MeV.
}
\end{figure}

In order to study the influence of the ISB forces on nuclear binding energies, we have performed the isocranking
calculations for the  $A$=34 isospin triplet. In these test calculations, the Coulomb interaction
has been switched off. The results with the CIB force (\ref{eq:class2}) are shown in Fig.~\ref{fig.16}.
As anticipated, this force  impacts the binding energy of $T_z=0$ system without
affecting the binding energies of $T_z=\pm 1$ triplet members. Hence, it changes the TDE indicator without
affecting  MDE. On the other hand, as shown in
Fig.~\ref{fig.17}, the class III force (\ref{eq:class3}) changes  MDEs without affecting  TDEs. This implies that the simplest strategy it to adjust the coupling constants $t_0^{{\rm II}}$ and $t_0^{{\rm III}}$
to TDEs and MDEs, respectively.

\begin{figure}[htb]
\begin{center}
\includegraphics[width=0.8\columnwidth]{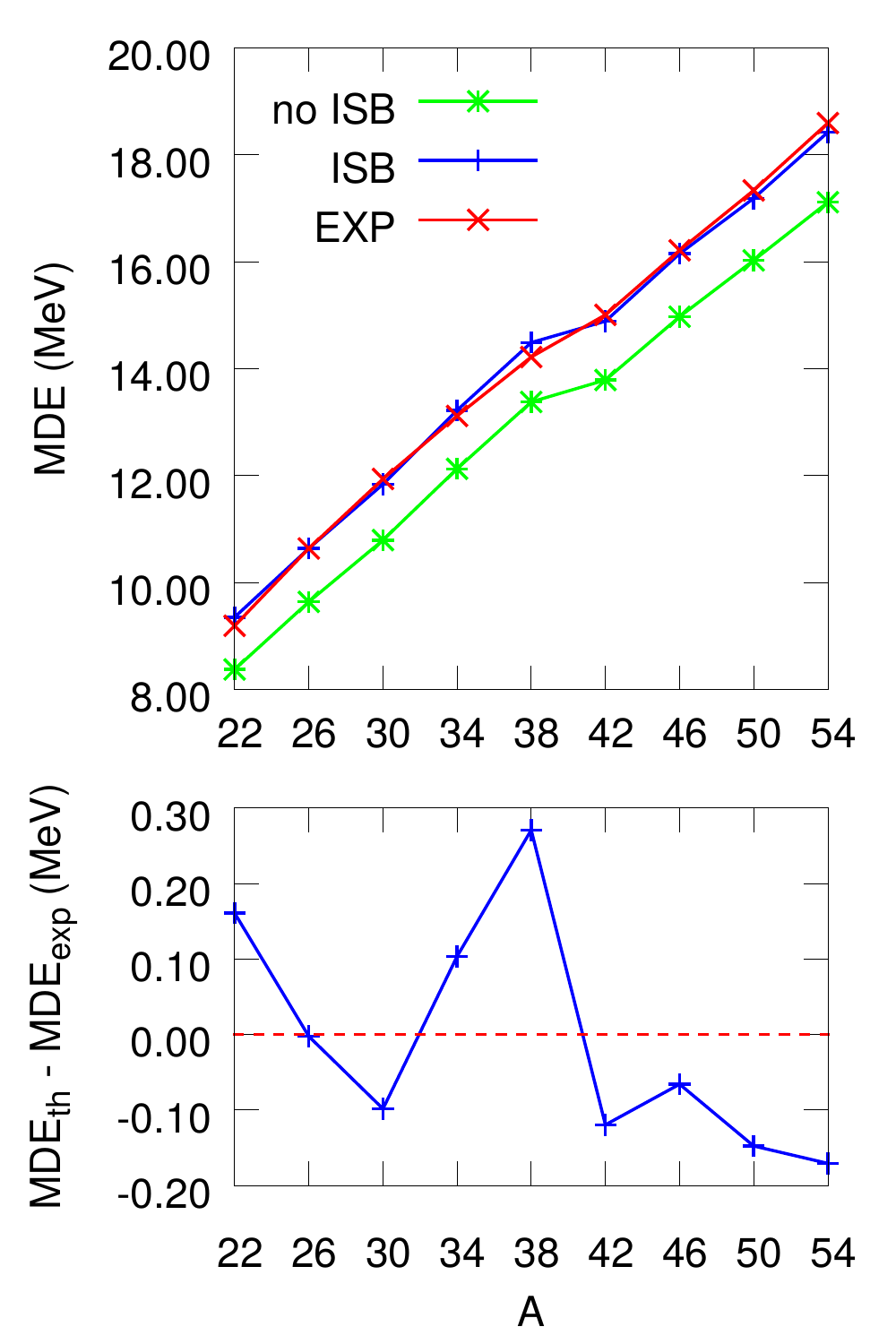}
\end{center}
\caption{\label{fig.18} Top: experimental and calculated (with and without hadronic ISB terms)  MDEs in the isospin triplets ranging from $A$=22 to 55. Bottom:  MDE residuals pertaining to ISB calculations.
}
\end{figure}

\begin{figure}[htb]
\begin{center}
\includegraphics[width=0.8\columnwidth]{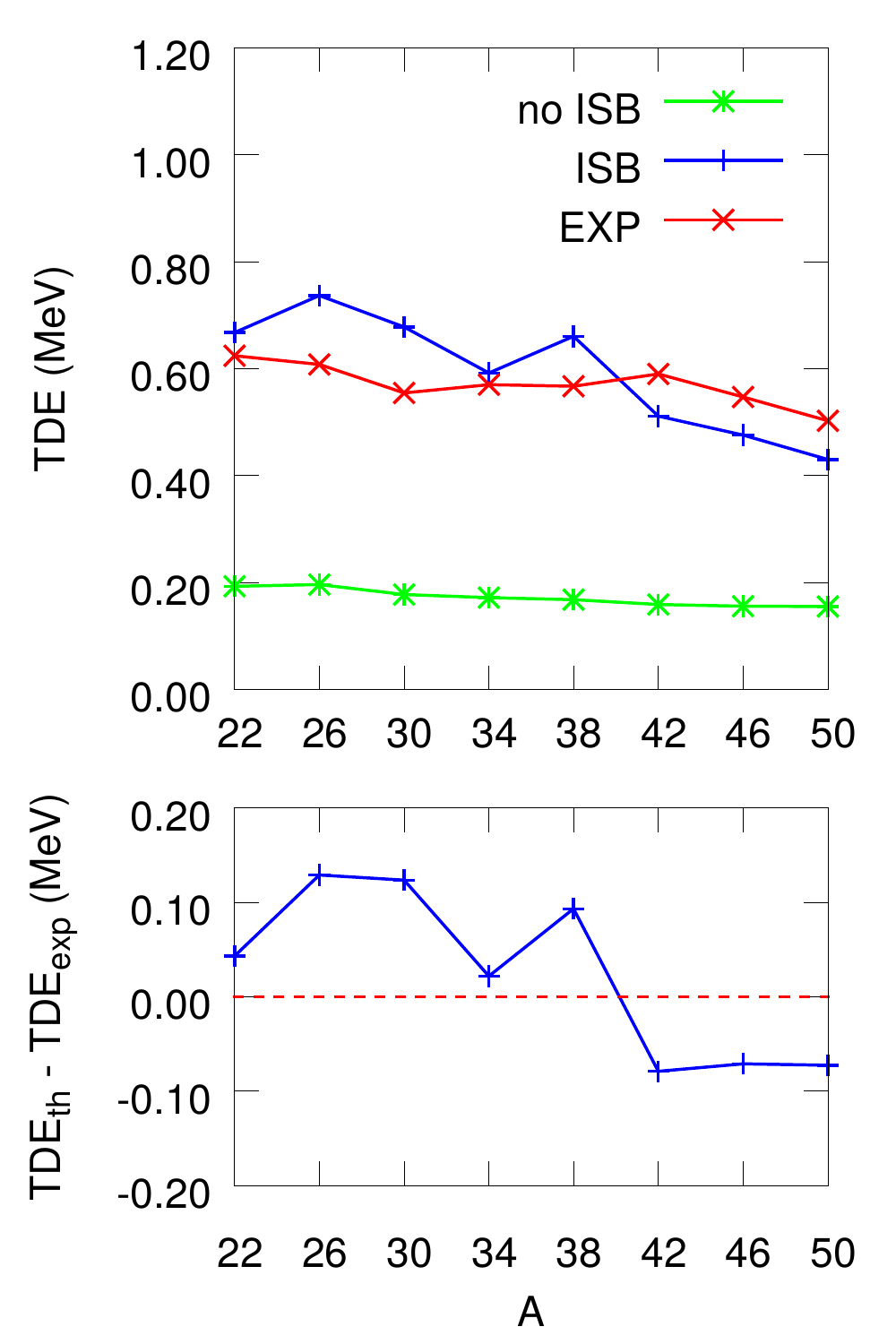}
\end{center}
\caption{\label{fig.19} Similar to Fig.~\ref{fig.18} but for TDEs.
}
\end{figure}

Figures~\ref{fig.18} and \ref{fig.19} show preliminary results of the calculated MDEs and TDEs in
the isospin triplets ranging from $A$=22 to 55.
The calculations have been done using the SV Skyrme EDF with Coulomb, augmented by the CIB and CSB forces with $t_0^{{\rm II}}=20$\,MeV and $t_0^{{\rm III}}=-8$\,MeV, respectively. It is seen that a generalized $pn$-mixed Skyrme DFT approach is able to reproduce experimental data on TEDs and MEDs.

The $pn$-mixed ISB DFT formalism  can address the evolution of ISB effects
with angular momentum in  rotational bands of $T=\frac12$ and $T=1$ nuclei.
Such effects has been investigated  within shell-model framework~\cite{[Zuk02],[Ben07a],[Dav13],[Kan12a],[Kan13],[Kan14],[Hen14]}.
The  quantities of interest are mirror energy differences (MEDs) and triplet energy
differences (TEDs),  similar to  MDEs and TDEs
but defined through excitation energies $\Delta E_I$ of rotational states at a given angular momentum  $I$:
\begin{eqnarray}
\mathrm{MED}(I) = \Delta E_I \left(T=\frac12,T_z=-\frac12\right)&\,& \nonumber \\
                 - \Delta E_I\left(T=\frac12,T_z=+\frac12\right), &\,&
\end{eqnarray}
\begin{eqnarray}
\mathrm{TED}(I) = \Delta E_I\left(T=1,T_z=-1\right)&\,& \nonumber \\
               - 2\Delta E_I\left(T=1,T_z=0\right)&\,& \nonumber \\
                 +\Delta E\left(T=1,T_z=+1\right). &\,&
\end{eqnarray}
In general, while shell-model studies indicate that the ISB hadronic forces are as important as the Coulomb interaction for the explanation of angular momentum dependence of MEDs and TEDs, the overall picture is not fully understood, especially in the upper $fp$ shell.

One of the key features of the NCCI model based on Skyrme forces with low effective mass is its  ability to
reproduce rotational bands. The preliminary calculations performed for  $^{48}$Cr, a key nucleus in the $f$-shell concerning collective properties, indicate that our model is capable of reproducing
its collective g.s. band fairly well~\cite{[Bac15]}. This opens  an opportunity
to address MED and TED effects in a non-perturbative way, as a function of angular momentum,  incorporating the ISB effects self-consistently along the rotational path.


\section{Proton-neutron pairing}
\label{Sect.06}

Superfluidity and superconductivity belong to the most spectacular examples of emergent phenomena in many-body systems.  They appear at different physical scales and in different environments in
atomic, condensed matter, nuclear, and elementary particle physics. The BCS mechanism~\cite{[Bar57]} behind the pair condensate
works irrespective of details of the underlying  interaction that couples fermions into the Cooper pairs at the Fermi surface. In low-energy nuclear physics, nucleonic pairing affects many properties of finite nuclei and nucleonic matter \cite{[Boh58],[BroZel]}.

Nucleonic Cooper pairs can exist in many flavors. In terms of isospin quantum number, one can talk about isovector triplet $(T=1)$ and isovector singlet $(T=0)$ pairs. The  conventional $nn$ and $pp$ pairing has isovector character, while the $pn$ pairing can be either isovector or isoscalar.
The first
attempts to incorporate the $pn$-pairing into the independent quasi-particle approach date back to the
mid-sixties~\cite{[Gos64],[Gos65],[Cam65]}.  These models were further developed into a consistent formalism
allowing for simultaneous treatment of isovector and isoscalar pairs~\cite{[Che66],[Che66a],[Che67],[Goo68],[Goo70]}.
The formalism was further extended to include stretched $T$=0 pairs ($\alpha\alpha$ $pn$-pairing) in Ref.~\cite{[Goo72]}.
These early approaches encountered difficulties in predicting coexisting
$T=0$ and $T=1$  pairing-phases. The solutions of the early models based on  separable seniority-type interactions could be conveniently classified in terms of  a single parameter, the value of isoscalar-to-isovector matrix element ratio
$x\equiv G^{T=0}/G^{T=1}$.
In particular, for $N=Z$ nuclei, the solution is of a pure isovector type for  $x<1$; for $x=1$ the isovector and isoscalar phases
are degenerate;  and for $x>1$ there appears a pure isoscalar solution. These models were  later extended~\cite{[Sat97a]} to include particle-number
projected wave functions. For $x> x_{\rm crit}\approx 1.1$, the extended models
have produced coexistence of $T=0$ and $T=1$ phases. It was also predicted that
for $x\geq 1.3$  the isoscalar pairing component could be the source of  the Wigner energy~\cite{[Sat97a],[Sat97]}. Following these developments, various models have been proposed to look for signatures of the isoscalar pairing phase and explain the Wigner energy~\cite{[Ner09],[San12],[Ben13a],[Car14],[Neg14],[Ben14a],[Fra14],[Sam15]}.

In spite of many efforts, however,  a comprehensive theory of nucleonic  pairing still eludes us. This negatively impacts our understanding of nuclei  in  the vicinity of the $N=Z$ line, where the $pn$-correlations
are expected to be strongest.  Indeed, a consistent approach to the the $pn$-pairing problem
requires implementing the $pn$-mixing on the mean-field level, whereby the s.p. wave functions are  combinations
of proton and neutron components. Basic self-consistency principles require such a mixing to accompany any hypothetical $pn$-mixing in the pairing channel. Moreover, the stability and existence of the $pn$-pairing condensate may critically depend on the restoring force related to the $pn$-mixing on the mean field level, and thus both must be simultaneously included in the theoretical description. As discussed above, the  DFT frameworks incorporating the  $pn$-mixing have been developed  recently~\cite{[Sat13c],[She14]}. This constitutes   the first step towards building a consistent symmetry-unrestricted superfluid DFT approach with the complete  $pn$-mixing. The experience gathered so far indicates that in order to capture structure of $N\approx Z$ nuclei, the $pn$-mixed formalism should be further extended to include the restoration of number symmetry and proper treatment of isospin~\cite{[Sat01],[Sat10]}.

\begin{figure}[htb]
\begin{center}
\includegraphics[width=0.8\columnwidth]{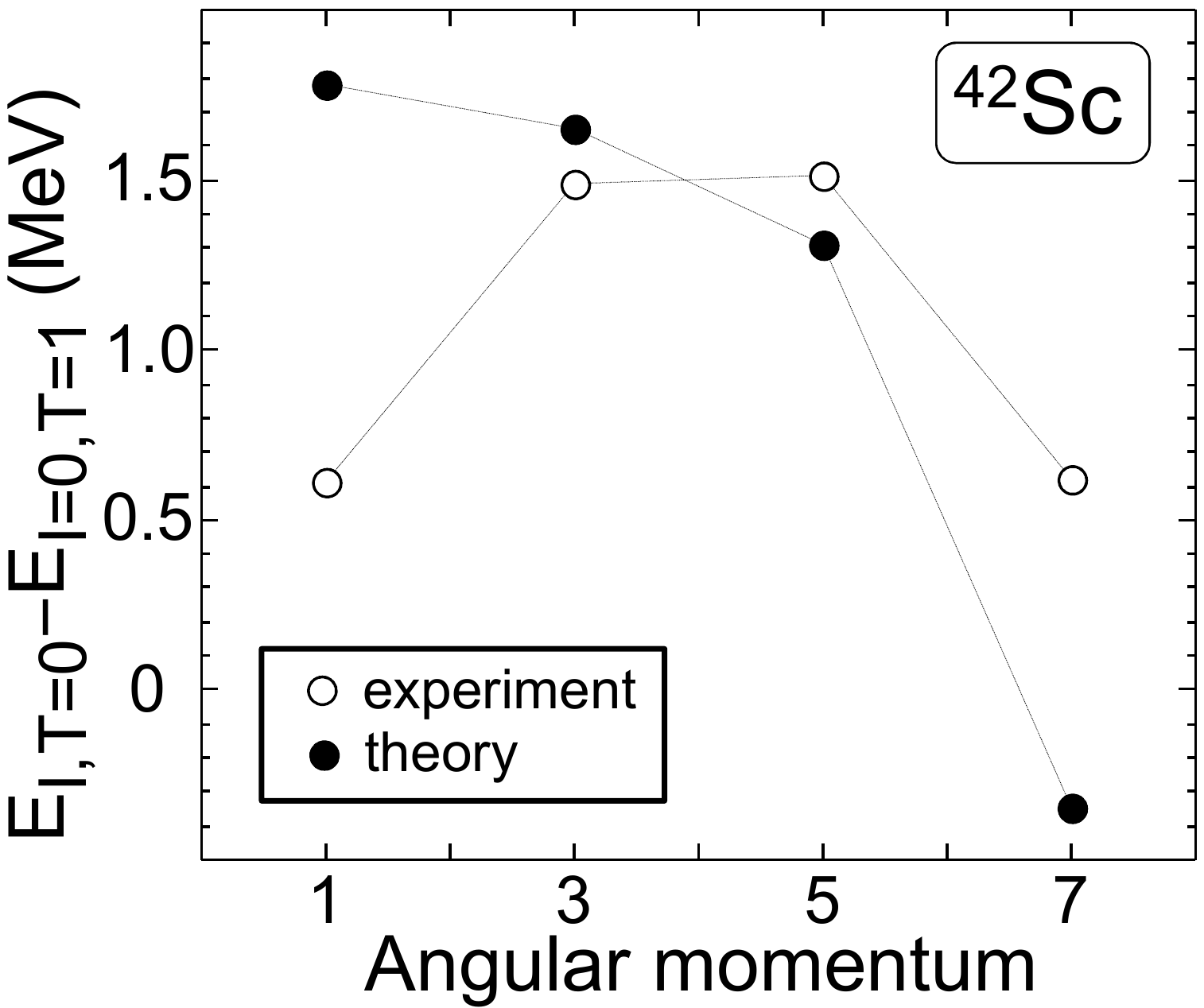}
\end{center}
\caption{\label{fig.20} Experimental (circles) and calculated (NCCI; dots) excitation energies of the $T=0$ states in $^{42}$Sc relative to the $0^+$ ground state.
}
\end{figure}
A need for the enhancement of the deuteron-like $J=1^+, T=0$ pairing channel is  apparent  in the present Skyrme-NCCI approach, which
accounts relatively well for low-spin states but has systematic problems with  the $J=1, T=0$ matrix elements in $N=Z$ nuclei.
This is illustrated in Fig.~\ref{fig.20} showing the calculated and experimental $T=0$ multiplet in
$^{42}$Sc. It is seen that the antialigned state $J=1^+, T=0$  is strongly underbound
and  the stretched $I=7^+$ level is overbound.  A similar situation has been encountered  in $^{6}$Li where the NCCI model  strongly underbinds $J=1, T=0$ matrix elements~\cite{[Sat15b]}.


\section{Prospects}
\label{Sect.07}

In this paper, we  overview selected topics related to isobaric spin  in $N\sim Z$ nuclei. We demonstrate that ``no-core" and self-consistency aspects of nuclear DFT are essential for our quantitative understanding of isospin breaking. The approximate nature of isospin  poses many theoretical challenges  as the underlying symmetry is broken both explicitly and spontaneously, and the
isospin mixing is highly non-perturbative due to a mismatch between the ranges of electrostatic and strong forces. To handle this situation theoretically, one needs to go beyond the usual single-reference DFT formalism.
The proposed multi-reference DFT framework, based on isospin- and angular-momentum projected wave functions, provides a very satisfactory description of isospin breaking effects in beta decay.

The future developments will undoubtedly utilize the newly developed isospin-invariant density functional framework~\cite{[Sat13c],[She14]}. The $pn$-mixed  DFT formalism needs to be extended to the particle-particle channel   by including pairing interaction of both isoscalar and isovector types.  This will enable us to study the importance of the isoscalar pairing densities and fields on the structure of $N\approx Z$ nuclei and the impact of $pn$-mixing on single- and double-beta decays. It seems, however,
that in order to capture key structural aspects of $N=Z$ nuclei, the $pn$-mixed formalism should be further extended to include the restoration of particle number, isospin, and angular momentum. This will require an appropriate NCCI extension.

In quantitative calculations, quality input is crucial. A construction and optimization  of a realistic Hamiltonian-based EDF is among the most urgent needs. This development  will not only improve the predictive power of the model but  will also help  addressing the burning question pertaining to the role of  ISB interaction components.


\begin{ack}
Useful discussions with Jacek Dobaczewski  are gratefully acknowledged. Pawe{\l} B{\c a}czyk is thanked for providing material
presented in Sec.~\ref{Sect.05} prior to publication \cite{[Bac15]}.
This article is
based upon work supported by the Polish National Science Centre (NCN) under Contract No. 2012/07/B/ST2/03907 and by the U.S. Department of Energy, Office of
Science, Office of Nuclear Physics under Award Numbers No.\
DOE-DE-SC0013365 (Michigan State University) and No.\ DE-SC0008511
(NUCLEI SciDAC Collaboration).
The CSC$-$IT Center for Science Ltd, Finland, is acknowledged for the allocation of computational resources.
\end{ack}




\begin{thebibliography}{100}

\bibitem{[Hei32a]}
{W. Heisenberg, Z. Phys. {\bf 77}, 1 (1932)}.

\bibitem{[Wig37]}
{E.P. Wigner, Phys. Rev. {\bf 51}, 106 (1937)}.

\bibitem{[Rob66]}
{D. Robson, Annu. Rev. Nucl. Sci. {\bf 16}, 119 (1966).}

\bibitem{[Wil69a]}
{D.H. Wilkinson, ed. {\it Isospin in nuclear physics}, \ (North Holland,
  Amsterdam, 1969)}.

\bibitem{[Har15]}
{J.C. Hardy and I.S. Towner, Phys. Rev. C {\bf 91}, 025501 (2015).}

\bibitem{[Hin12]}
{C.B. Hinke {\it et al.\/}, Nature {\bf 486}, 341 (2012).}

\bibitem{[Lid06]}
{S.N. Liddick {\it et al.\/}, Phys. Rev. Lett. {\bf 97}, 082501 (2006).}

\bibitem{[Sat97]}
{W. Satu{\l}a, D.J. Dean, J. Gary, S. Mizutori, and W. Nazarewicz, Phys. Lett.
  {\bf 407B}, 103 (1997)}.

\bibitem{[Fra14]}
{S. Frauendorf and A.O. Macchiavelli, Prog. Part. Nucl. Phys. {\bf 78}, 24-90
  (2013)}.

\bibitem{[Nak94a]}
{T. Nakatsukasa, K. Matsuyanagi, I. Hamamoto, and W. Nazarewicz, Nucl. Phys.
  {\bf A573}, 333 (1994).}

\bibitem{[Aue83]}
{N. Auerbach, Phys. Rep. {\bf 98}, 273 (1983)}.

\bibitem{[Sli65]}
{L.A. Sliv and Yu.I. Kharitonov, Phys. Lett. {\bf 16}, 176 (1965)}.

\bibitem{[Boh67]}
{A. Bohr, J. Damg{\aa}rd, and B. Mottelson, in {\it Nuclear Structure}, ed. by
  A. Hossian {\it et al.} (North-Holland Publ. Co., Amsterdam, 1967).}

\bibitem{[Cha97w]}
{E. Chabanat, {\it et al.\/}, Nucl. Phys. {\bf A627} (1997) 710; {\bf A635}
  (1998) 231.}

\bibitem{[Far03]}
{E. Farnea {\it et al.}, Phys. Lett. {\bf B551}, 56 (2003)}.

\bibitem{[Cor11x]}
{A. Corsi {\it et al.}, Acta Phys. Pol. {\bf B42}, 619 (2011); Phys. Rev. C
  {\bf 84}, 041304 (2011)}.

\bibitem{[Sat13]}
{W. Satu{\l}a, J. Dobaczewski, M. Konieczka, and W. Nazarewicz,
  arXiv:1307:1550}.

\bibitem{[Eng70]}
{C.A. Engelbrecht and R.H. Lemmer, Phys. Rev. Lett. {\bf 24}, 607 (1970)}.

\bibitem{[Cau80]}
{E. Caurier, A. Poves, and A. Zucker, Phys. Lett. B {\bf 96}, 11 (1980); {\bf
  96}, 15 (1980)}.

\bibitem{[Cau82]}
{E. Caurier and A. Poves, Nucl. Phys. {\bf A385}, 407 (1982)}.

\bibitem{[Sat11d]}
{W. Satu{\l}a, J. Dobaczewski, W. Nazarewicz, and M. Rafalski, Acta Phys.
  Polon. B {\bf 42}, 415 (2011)}.

\bibitem{[Bei75]}
{M. Beiner, H. Flocard, N. Van Giai, and P. Quentin, Nucl. Phys. {\bf A238}, 29
  (1975)}.

\bibitem{[Dob84]}
{J. Dobaczewski, H. Flocard, and J. Treiner, Nucl. Phys. A {\bf 422}, 103
  (1984)}.

\bibitem{[Sat10]}
{W. Satu{\l}a, J. Dobaczewski, W. Nazarewicz, and M. Rafalski, Phys. Rev. C
  {\bf 81}, 054310 (2010)}.

\bibitem{[Sat09a]}
{W. Satu{\l}a, J. Dobaczewski, W. Nazarewicz, and M. Rafalski, Phys. Rev. Lett.
  {\bf 103}, 012502 (2009)}.

\bibitem{[RS80]}
{P. Ring and P. Schuck, {\sl The Nuclear Many-Body Problem} (Springer-Verlag,
  Berlin, 1980)}.

\bibitem{[Dob09d]}
{J. Dobaczewski, W. Satu{\l}a, B.G. Carlsson, J. Engel, P. Olbratowski, P.
  Powa{\l}owski, M. Sadziak, J. Sarich, N. Schunck, A. Staszczak, M.V.
  Stoitsov, M. Zalewski, and H. Zdu\'nczuk, Comput. Phys. Commun. {\bf 180},
  2361 (2009)}.

\bibitem{[Zdu07a]}
{H. Zdu{\'n}czuk, W. Satu{\l}a, J. Dobaczewski, and M. Kosmulski, Phys. Rev. C
  {\bf 76}, 044304 (2007)}.

\bibitem{[Sat12]}
{W. Satu{\l}a, J. Dobaczewski, W. Nazarewicz, and T.R. Werner, Phys. Rev. C
  {\bf 86}, 054316 (2012)}.

\bibitem{[Sat11b]}
{W. Satu{\l}a, J. Dobaczewski, W. Nazarewicz, M. Borucki, and M. Rafalski, Int.
  J. Mod. Phys. {\bf E20}, 244 (2011)}.

\bibitem{[Ben03]}
{M. Bender, P.-H. Heenen, and P.-G. Reinhard, Rev. Mod. Phys. {\bf 75}, 121
  (2003)}.

\bibitem{[Lac09]}
{D. Lacroix, T. Duguet, and M. Bender, Phys. Rev. C {\bf 79}, 044318 (2009)}.

\bibitem{[Sat14b]}
{W. Satu{\l}a and J. Dobaczewski, Phys. Rev. C {\bf 90}, 054303 (2014).}

\bibitem{[Sad13a]}
{J. Sadoudi, M. Bender, K. Bennaceur, D. Davesne, R. Jodon, and T. Duguet,
  Phys. Scr. {\bf T154}, 014013 (2013)}.

\bibitem{[Bal14b]}
{B. Bally, B. Avez, M. Bender, and P.-H. Heenen, Phys. Rev. Lett. {\bf 113},
  162501 (2014).}

\bibitem{[Sat14]}
{W. Satu{\l}a, J. Dobaczewski, M. Konieczka, and W. Nazarewicz, Acta Phys.
  Polonica B{\bf 45}, 167 (2014).}

\bibitem{[Sat15]}
{W. Satu\l{}a, J. Dobaczewski, and M. Konieczka, JPS Conf. Proc. {\bf 6},
  020015 (2015).}

\bibitem{[Naz14]}
{W. Nazarewicz, P.-G. Reinhardt, W. Satu{\l}a, and D. Vretenar, Eur. Phys. J. A
  {\bf 50}, 20 (2014)}.

\bibitem{[Mar06b]}
{W.J. Marciano and A. Sirlin, Phys. Rev. Lett. {\bf 96}, 032002 (2006).}

\bibitem{[Tow08]}
{I.S. Towner and J.C. Hardy, Phys. Rev. C {\bf 77}, 025501 (2008)}.

\bibitem{[Tow94]}
{I.S. Towner, Phys. Lett. {\bf B333}, 13 (1994)}.

\bibitem{[Har09]}
{J.C. Hardy and I.S. Towner, Phys. Rev. C {\bf 79}, 055502 (2009)}.

\bibitem{[Tow10]}
{I.S. Towner and J.C. Hardy, Phys. Rev. C {\bf 82}, 065501 (2010)}.

\bibitem{[Tow10a]}
{I.S. Towner and J.C. Hardy, Rep. Prog. Phys. {\bf 73}, 046301 (2010)}.

\bibitem{[Har14]}
{J.C. Hardy and I.S. Towner, J. Phys. G: Nucl. Part. Phys. {\bf 41}, 114004
  (2014).}

\bibitem{[Cab63]}
{N. Cabibbo, Phys. Rev. Lett. {\bf 10}, 531 (1963)}.

\bibitem{[Kob73]}
{M. Kobayashi and T. Maskawa, Prog. Theor. Phys. {\bf 49}, 652 (1973)}.

\bibitem{[Oli14]}
{K.A. Olive {\it et al.\/} (Particle Data Group), Chin. Phys. C {\bf 38},
  090001 (2014).}

\bibitem{[Par14]}
{H.I. Park, J.C. Hardy, V.E. Iacob, M. Bencomo, L. Chen, V. Horvat, N. Nica, B.
  T. Roeder, E. Simmons, R.E. Tribble, and I. S. Towner, Phys. Rev. Lett. {\bf
  112}, 102502 (2014).}

\bibitem{[Bla14]}
{B. Blank {\it et al.\/}, Eur. Phys. J. A {\bf 51}, 8 (2014)}.

\bibitem{[Laf15]}
{A.T. Laffoley {\it et al.\/}, Phys. Rev. C {\bf 92}, 025502 (2015)}.

\bibitem{[Dam69b]}
{J. Damg{\aa}rd, Nucl. Phys. A{\bf 130}, 233, (1969)}.

\bibitem{[Orm95a]}
{W.E. Ormand and B.A. Brown, Phys. Rev. C {\bf 52}, 2455 (1995)}.

\bibitem{[Sag96a]}
{H. Sagawa, N. Van Gai, and T. Suzuki, Phys. Rev. C {\bf 53}, 2163 (1996)}.

\bibitem{[Lia09]}
{H. Liang, N. Van Giai, and J. Meng, Phys. Rev. C {\bf 79}, 064316 (2009)}.

\bibitem{[Aue09]}
{N. Auerbach, Phys. Rev. C {\bf 78}, 035502 (2009)}.

\bibitem{[Sat11a]}
{W. Satu{\l}a, J. Dobaczewski, W. Nazarewicz, and M. Rafalski, Phys. Rev. Lett.
  {\bf 106}, 132502 (2011)}.

\bibitem{[Mil08]}
{G.A. Miller and A. Schwenk, Phys. Rev. C {\bf 78}, 035501 (2008)}.

\bibitem{[Mil09]}
{G.A. Miller and A. Schwenk, Phys. Rev. C {\bf 80}, 064319 (2009)}.

\bibitem{[Sat15b]}
{W. Satu\l{}a, J. Dobaczewski, and M. Konieczka, in preparation}.

\bibitem{[Nak10]}
{K. Nakamura {\it et al.} (Particle Data Group), J. Phys. G {\bf 37}, 075021
  (2010)}.

\bibitem{[Poc04]}
{D. Pocanic {\it et al.}, Phys. Rev. Lett. {\bf 93}, 181803 (2004)}.

\bibitem{[Nav09a]}
{O. Naviliat-Cuncic and N. Severijns, Phys. Rev. Lett. {\bf 102}, 142302
  (2009)}.

\bibitem{[Cho93]}
{W.-T. Chou, E.K. Warburton, and B.A. Brown, Phys. Rev. C {\bf 47}, 163
  (1993)}.

\bibitem{[Mar96]}
{G. Martinez-Pinedo, A. Poves, E. Caurier, and A.P. Zucker, Phys. Rev. C {\bf
  53}, R2602 (1996).}

\bibitem{[Cau12]}
{E. Caurier, F. Nowacki, and A. Poves, Phys. Lett. B {\bf 711}, 62 (2012).}

\bibitem{[Hor13]}
{M. Horoi and B. A. Brown, Phys. Rev. Lett. {\bf 110}, 222502 (2013).}

\bibitem{[Vai09]}
{S. Vaintraub, N. Barnea, and D. Gazit, Phys. Rev. C {\bf 79}, 065501 (2009)}.

\bibitem{[Men11]}
{J. Men\'endez, D. Gazit, and A. Schwenk, Phys. Rev. Lett. {\bf 107}, 062501
  (2011).}

\bibitem{[Eng14]}
{J. Engel, F. \v{S}imkovic, and P. Vogel, Phys. Rev. C {\bf 89}, 064308
  (2014)}.

\bibitem{[Hol13]}
{J.D. Holt and J. Engel, Phys. Rev. C {\bf 87}, 064315 (2013).}

\bibitem{[Yao15]}
{J.M. Yao, L.S. Song, K. Hagino, P. Ring, and J. Meng, Phys. Rev. C {\bf 91},
  024316 (2015).}

\bibitem{[Klo13]}
{P. Klos, J. Men\'endez, D. Gazit, and A. Schwenk, Phys. Rev. D {\bf 88},
  083516 (2013)}.

\bibitem{[Eks14]}
{A. Ekstr{\"o}m, G.R. Jansen, K.A. Wendt, G. Hagen, T. Papenbrock, S. Bacca, B.
  Carlsson, and D. Gazit, Phys. Rev. Lett. {\bf 113}, 262504 (2014).}

\bibitem{[Yak05]}
{K. Yako {\it et al.\/}, Phys. Lett. {\bf B615}, 193 (2005)}.

\bibitem{[Sas09]}
{M. Sasano {\it et al.\/}, Phys. Rev. C {\bf 79}, 024602 (2009)}.

\bibitem{[Kon15]}
{M. Konieczka, P. B{\c a}czyk, and W. Satu{\l}a, arXiv:1509.0448}.

\bibitem{[Kne12]}
{A. Knecht {\it et al.\/}, Phys. Rev. Lett. {\bf 108}, 122502 (2012)}.

\bibitem{[Sch12]}
{N. Schunck, J. Dobaczewski, J. McDonnell, W. Satu{\l}a, J.A. Sheikh, A.
  Staszczak, M. Stoitsov, and P. Toivanen, Comput. Phys. Commun. {\bf 183}, 166
  (2012)}.

\bibitem{[Bro85]}
{B.A. Brown and B.H. Wildenthal, At. Data Nucl. Data Tables {\bf 33}, 347
  (1985)}.

\bibitem{[Sek87]}
{T. Sekine, J. Cerny, R. Kirchner, O. Klepper, V.T. Koslowsky, A. P{\l}ochocki,
  E. Roeckl, D. Schardt, and B. Sherrill, Nucl. Phys. {\bf A467} (1987) 93}.

\bibitem{[Ike64]}
{K. Ikeda, Prog. Theor. Phys. {\bf 31}, 434 (1964)}.

\bibitem{[Sev08]}
{N. Severijns, M. Tandecki, T. Phalet, and I.S. Towner, Phys. Rev. C {\bf 78},
  055501 (2008)}.

\bibitem{[Nav09b]}
{O. Naviliat-Cuncic and N. Severijns, Eur. Phys. J. A {\bf 42}, 327 (2009)}.

\bibitem{[Mac01a]}
{R. Machleidt, Phys. Rev. C {\bf 63}, 024001 (2001)}.

\bibitem{[Bor15]}
{Sz. Borsanyi, S. Durr, Z. Fodor, C. Hoelbling, S.D. Katz, S. Krieg, L.
  Lellouch, T. Lippert, A. Portelli, K.K Szabo, and B.C. Toth, Science {\bf
  347}, 1452 (2015)}.

\bibitem{[Nol69]}
{J.A. Nolen and J.P. Schiffer, Ann. Rev. Nuc. Sci. {\bf 19}, 471 (1969)}.

\bibitem{[Kan14]}
{K. Kaneko, Y. Sun, T. Mizusaki, and S. Tazaki, Phys. Rev. C {\bf 89},
  031302(R) (2014).}

\bibitem{[Tu14]}
{X.L. Tu, Y. Sun, Y.H. Zhang, H.S. Xu, K. Kaneko, Yu.A. Litvinov, and M. Wang,
  J. Phys. G {\bf 41}, 025104 (2014)}.

\bibitem{[Ben15]}
{M.A. Bentley, S.M. Lenzi, S.A. Simpson, and C. Aa. Diget, Phys. Rev. C{\bf
  92}, 024310 (2015)}.

\bibitem{[Mil95]}
{G.A. Miller, and W.H.T. van Oers, \textit{Symmetries and Fundamental
  Interactions in Nuclei} edited by W.C. Haxton and E.M. Henley (World
  Scientific, Singapore, 1995).}

\bibitem{[Hen79]}
{E.M. Henley, and G.A. Miller, in \textit{Mesons in Nuclei}, edited by M. Rho
  and D.H. Wilkinson (North Holland, Amsterdam, 1979), p. 405.}

\bibitem{[Wir13]}
{R.B. Wiringa, S. Pastore, S.C. Pieper, and G.A. Miller, Phys. Rev. C {\bf 88},
  044333 (2013).}

\bibitem{[Orm89a]}
{W.E. Ormand and B.A. Brown, Nucl. Phys. A{\bf 491}, 1 (1989}.

\bibitem{[Zuk02]}
{A.P. Zuker, S.M. Lenzi, G. Martinez-Pinedo, and A. Poves, Phys. Rev. Lett.
  {\bf 89}, 142502 (2002)}.

\bibitem{[Ben07a]}
{M.A. Bentley and S.M. Lenzi, Prog. Part. Nucl. Phys. {\bf 59}, 497 (2007)}.

\bibitem{[Qi08]}
{C. Qi and F.R. Xu, Nucl. Phys. A{\bf 814}, 48 (2008)}.

\bibitem{[Kan12a]}
{K. Kaneko, T. Mizusaki, Y. Sun, S. Tazaki, and G. de Angelis, Phys. Rev. Lett.
  {\bf 109}, 092504 (2012).}

\bibitem{[Kan13]}
{K. Kaneko, Y. Sun, T. Mizusaki, and S. Tazaki, Phys. Rev. Lett. {\bf 110},
  172505 (2013).}

\bibitem{[Lam13]}
{Yi Hua Lam, N.A. Smirnova, and E. Caurier, Phys. Rev. C {\bf 87}, 054304
  (2013)}.

\bibitem{[Sag95a]}
{H. Sagawa, N. Van Giai, and T. Suzuki, Phys. Lett. B{\bf 353}, 7 (1995).}

\bibitem{[Bro98a]}
{B.A. Brown, Phys. Rev. C {\bf 58}, 220 (1998)}.

\bibitem{[Bro00b]}
{B.A. Brown, W.A. Richter, and R. Lindsay, Phys. Lett. {\bf B483}, 49 (2000)}.

\bibitem{[Tho52]}
{R.G. Thomas, Phys. Rev. {\bf 88}, 1109 (1952)}.

\bibitem{[Ehr51]}
{J.B. Ehrman, Phys. Rev. {\bf 81}, 412 (1951)}.

\bibitem{[Gri02]}
{L.V. Grigorenko, I.G. Mukha, I.J. Thompson, and M.V. Zhukov, Phys. Rev. Lett.
  {\bf 88}, 042502 (2002)}.

\bibitem{[Mic10]}
{N. Michel, W. Nazarewicz, and M. P\l{}oszajczak, Phys. Rev. C {\bf 82}, 044315
  (2010)}.

\bibitem{[Yua14]}
{C. Yuan, C. Qi, F. Xu, T. Suzuki, and T. Otsuka, Phys. Rev. C {\bf 89}, 044327
  (2014)}.

\bibitem{[Per04]}
{E. Perli\'nska, S.G. Rohozi\'nski, J. Dobaczewski, and W. Nazarewicz, Phys.
  Rev. C {\bf 69}, 014316 (2004)}.

\bibitem{[Sat13c]}
{K. Sato, J. Dobaczewski, T. Nakatsukasa, and W. Satu{\l}a, Phys. Rev. C {\bf
  88}, 061301(R) (2013)}.

\bibitem{[She14]}
{J.A. Sheikh, N.~Hinohara, J.~Dobaczewski, T.~Nakatsukasa, W.~Nazarewicz, and
  K.~Sato, Phys. Rev. C {\bf 89}, 054317 (2014)}.

\bibitem{[Glo04w]}
{S. G{\l}owacz, W. Satu{\l}a, and R.A. Wyss, Eur. Phys. J. {\bf A19} (2004)
  33.}

\bibitem{[Sat01]}
{W. Satu{\l}a and R. Wyss. Phys. Rev. Lett. {\bf 86}, 4488 (2001); {\bf 87},
  052504 (2001)}.

\bibitem{[Dav13]}
{P.J. Davies {\it et al.\/}, Phys. Rev. Lett. {\bf 111}, 072501 (2013)}.

\bibitem{[Hen14]}
{J. Henderson {\it et al.\/}, Phys. Rev. C {\bf 90}, 051303 (2014)}.

\bibitem{[Bac15]}
{P. B{\c a}czyk, M. Konieczka, and W. Satu{\l}a, in preparation}.

\bibitem{[Bar57]}
{J. Bardeen, L.N. Cooper, and J.R. Schrieffer, Phys. Rev. {\bf 108}, 1175
  (1957)}.

\bibitem{[Boh58]}
{A. Bohr, B.R. Mottelson, and D. Pines, Phys. Rev. {\bf 110}, 936, (1958)}.

\bibitem{[BroZel]}
{R.A. Broglia and V. Zelevinsky, editors {\it Fifty Years of Nuclear BCS\/},
  World Scientific, (2012)}.

\bibitem{[Gos64]}
{A. Goswami, Nucl. Phys. {\bf 60}, 228 (1964)}.

\bibitem{[Gos65]}
{A. Goswami and L.S. Kisslinger, Phys. Rev. {\bf 140}, B26 (1965)}.

\bibitem{[Cam65]}
{P. Camiz, A. Covello, and M. Jean, Nouvo Cimento {\bf 36}, 663 (1965); {\bf
  42}, 199 (1966)}.

\bibitem{[Che66]}
{H.T. Chen, G.L. Struble, and A. Goswami, Nucl. Phys. {\bf 88}, 208 (1966)}.

\bibitem{[Che66a]}
{H.T. Chen and A. Goswami, Nucl. Phys. {\bf 88}, 208 (1966)}.

\bibitem{[Che67]}
{H.T. Chen and A. Goswami, Phys. Lett. {\bf B24}, 257 (1967)}.

\bibitem{[Goo68]}
{A.L. Goodman, G.L Struble, and A. Goswami, Phys. Lett. {\bf B26}, 260 (1968)}.

\bibitem{[Goo70]}
{A.L. Goodman, G.L Struble, J. Bar-Touv, and A. Goswami, Phys. Rev. {\bf C2},
  380 (1970)}.

\bibitem{[Goo72]}
{A.L. Goodman, Nucl. Phys. {\bf A186}, 475 (1972)}.

\bibitem{[Sat97a]}
{W. Satu{\l}a and R. Wyss, Phys. Lett. {\bf B393}, 1 (1997)}.

\bibitem{[Ner09]}
{K. Neergard, Phys. Rev. C{\bf 80}, 044313 (2009).}

\bibitem{[San12]}
{N. Sandulescu, D. Negrea, and C.W. Johnson, Phys. Rev. C {\bf 86}, 041302(R)
  (2012).}

\bibitem{[Ben13a]}
{I. Bentley and S. Frauendorf, Phys. Rev. C {\bf 88}, 014322 (2013).}

\bibitem{[Car14]}
{B.G. Carlsson and J. Toivanen, Phys. Rev. C {\bf 89}, 054324 (2014).}

\bibitem{[Neg14]}
{D. Negrea and N. Sandulescu, Phys. Rev. C {\bf 90}, 024322 (2014).}

\bibitem{[Ben14a]}
{I. Bentley, K. Neergard, and S. Frauendorf, Phys. Rev. C {\bf 89}, 034302
  (2014).}

\bibitem{[Sam15]}
{M. Sambarto, N. Sandulescu, and C.W. Johnson, Phys. Lett. {\bf B740}, 137
  (2015).}

\end{thebibliography}

\end{document}